\PassOptionsToPackage{pdfpagelabels=false}{hyperref} 
\documentclass[fleqn,usenatbib]{mnras}
\usepackage{newtxtext,newtxmath}
\usepackage[T1]{fontenc}
\usepackage{ae,aecompl}

\usepackage{textcmds}
\usepackage{booktabs}

\usepackage{graphicx}
\usepackage{braket}
\usepackage{amsmath}
\usepackage{times}

\newcommand{\yr}                         {\,{\rm yr}}
\newcommand{\peryr}                    {\,{\rm yr}^{-1}}

\newcommand{\Gyr}                      {\,{\rm Gyr}}

\newcommand{\pkpc}                      {\,{\rm{pkpc}}}

\newcommand{\ckpc}                      {\,{\rm ckpc}}
\newcommand{\cMpc}                      {\,{\rm cMpc}}
\newcommand{\Msun}                    {\,{\rm M}_\odot}

\newcommand{\Msunyrsqkpc}                  {\,{\rm M}_\odot\,{\rm yr}^{-1}\,{\rm kpc}^{-2}}

\newcommand{\cmcubed}              {{\rm cm}^{-3}}

\newcommand{\K}                          {\,{\rm K}}

\title[Morphology and alignment of star-forming gas]
{The morphology of star-forming gas and its alignment with galaxies and dark matter haloes in the EAGLE simulations} 

\author[A. D. Hill et al.]{
Alexander D. Hill,$^{1}$\thanks{E-mail: a.d.hill@2017.ljmu.ac.uk (ADH)} Robert A. Crain,$^{1}$ Juliana Kwan$^{1}$ and Ian G. McCarthy.$^{1}$
\\
$^{1}$Astrophysics Research Institute, Liverpool John Moores University, 146 Brownlow Hill, Liverpool L3 5RF, UK
}

\date{Accepted XXX. Received YYY; in original form ZZZ}

\pubyear{2021}

\begin{document}
\label{firstpage}
\pagerange{\pageref{firstpage}--\pageref{lastpage}}

\maketitle
\begin{abstract}
We present measurements of the morphology of star-forming gas in galaxies from the EAGLE simulations, and its alignment relative to stars and dark matter (DM). Imaging of such gas in the radio continuum enables weak lensing experiments that complement traditional optical approaches. Star-forming gas is typically more flattened than the stars and DM within halo centres, particularly for present-day structures of total mass $\sim 10^{12-12.5}\Msun$, which preferentially host star-forming galaxies with rotationally supported stellar discs. Such systems have oblate, spheroidal star-forming gas distributions, but in both less- and more-massive subhaloes the distributions tend to be prolate, and its morphology correlates positively and significantly with that of its host galaxy's stars, both in terms of sphericity and triaxiality. The minor axis of star-forming gas most commonly aligns with the minor axis of its host subhalo's central DM distribution, but this alignment is often poor in subhaloes with a prolate DM distribution. Star-forming gas aligns with the DM at the centre of its parent subhalo less strongly than is the case for stars, but its morphological minor axis aligns closely with its kinematic axis, affording a route to observational identification of the unsheared morphological axis. The projected ellipticities of star-forming gas in EAGLE are consistent with shapes inferred from high-fidelity radio continuum images, and they exhibit greater shape noise than is the case for images of the stars, owing to the greater characteristic flattening of star-forming gas with respect to stars.
\end{abstract}


\section{Introduction}

The currently preferred $\Lambda$-cold dark matter ($\Lambda$CDM) cosmogony posits that the large-scale cosmic matter distribution (spatial scales $\gtrsim 1$ Mpc) is best described as a highly non-uniform system of voids, sheets, filaments and haloes, colloquially termed the `cosmic web'. This structure forms in response to the gravitational growth of small instabilities in the matter distribution of the early Universe \citep[e.g.][]{bond, faucher, shandarin}. Spectroscopic redshift surveys have revealed that galaxies are themselves distributed in a cosmic web, as expected if they broadly trace the underlying matter distribution. The non-uniform distribution of galaxies was apparent in early redshift surveys \citep[see e.g.][]{delapparent86,geller89}, but was demonstrated spectacularly by those exploiting the advent of highly multiplexed spectrographs, notably the 2dF Galaxy Redshift Survey \citep[2dFGRS;][]{colless} and the Sloan Digital Sky Survey \citep[SDSS;][]{tegmark}. 

A fundamental tenet of galaxy formation models is that galaxies form within the dark matter (DM) haloes that permeate the cosmic web \citep[e.g.][]{white_rees_78}. The broad correspondence between the clustering of galaxies inferred from observational surveys on one hand, and on the other that of the galaxies that form in semi-analytic models of galaxy formation \citep[e.g.][]{kauffmann99,springel05_ms,wechsler06,guo11} and, more recently, hydrodynamical simulations of large cosmic volumes \citep[e.g.][]{crain17,mccarthy17, springel18}, can be considered a remarkable corroboration of the $\Lambda$CDM paradigm. However, being subject to the rich array of dissipative physical processes that govern their growth, galaxies inevitably represent imperfect tracers of their local environment \citep[e.g.][]{kaiser84,white87}, such that their baryonic components do not necessarily trace the shape and orientation of their DM haloes in a simple fashion. 

Besides their potential use as a means to place constraints on the ill-understood microphysics of galaxy formation, and to reveal the nature of the environment of galaxies \citep[e.g.][]{codis15, zhang}, differences in the shape and orientation of baryonic components of galaxies with respect to those of their DM haloes are of particular interest because they represent sources of uncertainty in observational inferences of the morphology of DM haloes, and of their orientation with respect to the large-scale matter distribution \citep[e.g.][]{troxel}. This is of consequence for efforts to constrain cosmological parameters via the shape correlation function of galaxies, a key aim of ongoing optical/near-infrared weak lensing surveys such as the Canada-France-Hawaii Telescope Lensing Survey \citep[CFHTLens;][]{erben13}, Kilo-Degree Survey \citep[KiDS;][]{dejong15}, the Hyper Suprime Cam Subaru Strategic Program \citep[HSC;][]{aihara} and the Dark Energy Survey \citep[DES;][]{des05}, and ambitious forthcoming surveys with the Vera Rubin Observatory \citep{lsst09}, the \textit{Euclid} spacecraft \citep{laureijs12}, and the \textit{Nancy Grace Roman Space Telescope} \citep[e.g.][]{spergel15}. Moreover, the severity of differences between the shape and alignment of haloes and those of the observable structures used to infer them, has a strong bearing on the accuracy of weak gravitational lensing predictions derived from dark matter-only simulations. At present, such simulations are the only means of modelling the evolution of cosmic volumes comparable to those mapped out by lensing surveys.

Simplified techniques such as halo occupation distribution (HOD) modelling, subhalo abundance matching (SHAM), and semi-analytic models have, in order of increasing sophistication, proven valuable means of understanding the connection between galaxies and the matter distribution \citep[see e.g.][]{schneider_bridle_10,joachimi_13}. However, such methods have been shown to exhibit significant systematic differences with respect to the predictions of cosmological hydrodynamical simulations on small-to-intermediate spatial scales \citep[e.g.][]{chavesmontero16,springel18}, in large part because they (by design) do not self-consistently capture the back-reaction of baryon evolution on the structure of DM haloes \citep{bett10,guo16}. A comprehensive understanding of the influence of systematic uncertainties stemming from the differences in the shape and orientation of galaxies and their host haloes therefore requires self-consistent and realistic physical models of galaxy formation in a fully cosmological framework. 

There is a rich history of the use of numerical simulations to establish the correspondence between the morphology, angular momentum and orientation of galaxies, their satellite systems and their host DM haloes, with particular emphases on the roles played by gas accretion \citep[e.g.][]{chen03,sharma_steinmetz_05,sales_12}, mergers \citep[e.g.][]{dubinski_98,mbk_06,naab_06} and environment \citep[e.g.][]{croft_09,hahn_10,shao16}. However, prior studies have tended to suffer from one or more significant shortcomings, namely relatively poor spatial and mass resolution, relatively small sample sizes, and a poor correspondence between the properties of simulated galaxies with observed counterparts. These shortcomings are significantly ameliorated by the current generation of state-of-the-art hydrodynamical simulations, such as EAGLE \citep{crain15, schaye15}, HorizonAGN \citep{dubois16}, Illustris/IllustrisTNG \citep[e.g.][]{vogelsberger14,pillepich18} and MassiveBlack-II \citep{khandai15}. Each of these simulations broadly reproduces key observed properties of the present-day galaxy population, thus engendering confidence that they capture (albeit with varying degrees of accuracy) the complexity of the interaction between the baryonic components of galaxies and their DM haloes. The simulations each follow a cosmological volume sufficient to a yield representative galaxy population ($\sim 100^3 \cMpc^3$), and do so with a mass resolution ($\sim 10^6\Msun$) and spatial resolution ($\sim 1\pkpc$) that enables examination of the properties and evolution of even sub-$L^\star$ galaxies. Moreover, they capture important second-order effects such as the back-reaction of baryons on the structure and clustering of DM haloes. 

The emergence of optical weak lensing surveys as a promising means of constraining the nature of DM and dark energy has intensified the need to assess the severity of systematic uncertainties afflicting cosmic shear measurements (specifically, the galaxy shape correlation function). Cosmological hydrodynamical simulations have proven a valuable tool for this purpose, highlighting that galaxies can be significantly misaligned with respect to their DM haloes \citep[e.g.][]{bett10,hahn_10,bett12,tenneti14,vel15a,shao16,chisari17} and that the shapes and alignments of galaxies and their haloes are correlated over large distances via tidal forces \citep[][]{tenneti14,chisari15,codis15,vel15b}. The simulations have also been exploited to examine the morphological and kinematic alignment of galaxies with the  cosmic large-scale structure \citep[see e.g.][]{cuesta,codis18}. The current generation of state-of-the-art simulations remains reliant on the use of subgrid treatments of many of the key physical processes governing galaxy evolution and, as noted by \citet{joachimi_review} the details of their particular implementation can in principle influence the alignment of cosmic structures \citep[see also][]{vel15a}. However, in key respects the simulations appear to be quantitatively compatible with extant observational constraints, e.g. the $w_{\mathrm{g+}}$ correlation function of luminous red galaxies in SDSS and their analogues in the MassiveBlack-II simulation \citep[][their Fig. 21]{tenneti15}.

A complementary approach to optical/near-IR weak lensing surveys is to measure shear at radio frequencies. The concept has been demonstrated both by exploiting very large area, low source density radio data \citep{chang04}, and deep, pointed observations with greater source density \citep{patel10}. Ambitious future radio continuum surveys such as those envisaged for the Square Kilometre Array (SKA) may prove to be competitive with the largest optical surveys. An SKA Phase-1 continuum survey of $5000$ deg$^2$ is predicted to observe a source density of resolved star-forming galaxies of 2.7 arcmin$^{-2}$ \citep{ska_red}. \citet{brown15} argue that, in the most optimistic case, a full Phase-2 SKA survey over 3$\pi$ steradians would yield twice the areal coverage of the \textit{Euclid} `wide survey', with a similar source density of $\simeq$30 galaxies arcmin$^{-2}$. 

The characteristic redshift of sensitive radio continuum surveys may also prove to be significantly greater than that of optical counterparts. By bridging the gap between traditional shear measurements and those derived from maps of the cosmic microwave background (CMB) radiation, radio weak lensing surveys offer the promise of tomographic mapping of cosmic structure evolution in both the quasi-linear and strongly non-linear regimes. Shear mapping in the radio regime offers advantageous complementarity with optical surveys, in particular to suppress key systematic uncertainties. For example, the use of kinematic and/or polarisation information may enable improved characterisation of the intrinsic (unsheared) ellipticity, and suppress the influence of intrinsic alignment, the deviation from random of the observed ellipticity of a sample \citep{blain, morales,deburghday,whittaker15}.

Shear measurements in the radio regime are derived from images of the extended radio continuum emission from galaxies, which effectively traces the star-forming component of the interstellar medium (ISM). The morphology and kinematics of this component, and their relationship with those of the underlying DM distribution, can in principle differ markedly from the analogous quantities traced by the stellar component imaged by conventional lensing surveys. However, by design, leading models of the radio continuum sky \citep[e.g.][]{wilman_08,bonaldi_19} do not account for such differences. This therefore motivates an extension of prior examinations of the relationship between galaxies and the overall matter distribution, and correlation of shapes and alignments of galaxies separated over cosmic distances, focusing on the use of the star-forming ISM to characterise the morphology and orientation of galaxies. The current generation of state-of-the-art cosmological hydrodynamical simulations are well suited to this application since, as for the stellar component, they self-consistently model the evolution of star-forming gas within galaxies, including cosmological accretion from the intergalactic and circumgalactic media (IGM and CGM, respectively), expulsion by feedback processes, and its interaction with a dynamically `live' DM halo. 

In this study, we use the cosmological hydrodynamical simulations of the EAGLE project \citep{schaye15,crain15} to examine the correspondence between the morphology and orientation of the star-forming ISM of galaxies and those of their parent DM haloes. EAGLE is well suited to this application: although the simulations do not explicitly model the balance between molecular, atomic and ionised hydrogen, the use of empirical or theoretical models to partition gas into these phases indicates that the simulations broadly reproduce key properties of the atomic and molecular reservoirs of galaxies \citep[see e.g.][]{lagos15,bahe16,crain17,dave20} including, crucially, the `fundamental plane of star formation' that relates their stellar mass, star formation rate and neutral hydrogen fraction \citep{lagos16}. This study complements prior examinations of the morphology of stars, hot gas and DM in the EAGLE simulations \citep[e.g.][]{vel15a,vel15b,shao16}. The morphology of the star-forming ISM of galaxies in the IllustrisTNG-50 simulation (hereafter TNG50) was also examined by \citet{pillepich19}; whilst the motivation for that study was quite different to that of ours, their findings are of direct relevance and offer an opportunity to assess the degree of consensus between different simulations. 

This paper is structured as follows. We discuss our numerical methods in Section \ref{sec:methods}, as well as summarising briefly details of the EAGLE simulation and galaxy finding algorithms, and our sample selection criteria. In Section \ref{sec:results_morphology} we examine the morphology of star-forming gas and its dependence on subhalo mass and redshift. In Section \ref{sec:results_alignments} we examine the internal alignment of star-forming gas with DM and stars, and its mutual alignment with its kinematic axis, again as a function of subhalo mass and redshift. In Section \ref{sec:results_2d} we investigate the shapes and alignments of the various matter components in 2D. In Section \ref{sec:Discussion} we discuss and summarise our findings. In a series of appendices, we examine the influence of a series of numerical and modelling factors on our findings. 

\section{Methods}
\label{sec:methods}

In this section we briefly introduce the EAGLE simulation (Section \ref{sec:Eagle}) and key numerical techniques for identifying haloes and subhaloes (Section \ref{sec:halo_def}), and for characterising their morphology with shape parameters (Section \ref{sec:redit}). Our sample selection criteria are discussed in Section \ref{sec:sample}. Detailed descriptions of the simulations are provided by many other studies using them, so we present only a concise summary of the most relevant aspects and refer the interested reader to the project's reference articles \citep{crain15, schaye15}. 

\subsection{Simulations}
\label{sec:Eagle}

The EAGLE project (the Evolution and Assembly of GaLaxies and their Environments) comprises a suite of hydrodynamical simulations that model the formation and evolution of galaxies and the cosmic large-scale structure in a $\Lambda$CDM cosmogony \citep{crain15, schaye15}. Particle data, and derived data products, from the simulations have been released to the community as detailed by \citet{mcalpine16}.
The simulations were evolved with a modified version of the Tree-Particle-Mesh (TreePM) smoothed particle hydrodynamics (SPH) solver \textsc{Gadget-3} \citep[last described by][]{springel05_gad}. The main modifications include the implementation of the pressure-entropy formulation of SPH introduced by \citet{hopkins}, a time-step limiter as proposed by \citet{durier12}, switches for artificial viscosity and artificial conduction, as per \citet{cullen10} and \citet{price08}, respectively, and the use of the \citet{wendland95} $C^2$ smoothing kernel. The influence of these developments on the properties of the galaxy population yielded by the simulations is explored by \citet{schaller15b}.

EAGLE includes subgrid treatments of several physical processes that are unresolved by the simulations. These include element-by-element radiative heating and cooling of 11 species \citep{WSS_A} in the presence of a spatially uniform, temporally evolving UV/X-ray background radiation field \citep{haardt01} and the cosmic microwave background (CMB); a model for the treatment of the multiphase ISM as a single-phase fluid with a polytropic pressure floor \citep{schaye_DV}; a metallicity-dependent density threshold above which gas becomes eligible for star formation \citep{schaye04}, with a probability of conversion dependent on the gas pressure \citep{schaye_DV}; stellar evolution and mass-loss \citep{WS_B}; the seeding of BHs and their growth via gas accretion and mergers \citep{springel05_agn, booth_schaye, rosas_guevara}; and feedback associated with the formation of stars \citep{dallavecchia} and the growth of BHs \citep{booth_schaye,schaye15}. The simulations adopt the stellar initial mass function (IMF) of \citet{chabrier03}. The efficiency of stellar feedback was calibrated to reproduce the stellar mass function of the low-redshift galaxy population and, broadly, the sizes of local disc galaxies. The efficiency of AGN feedback was calibrated to reproduce the present-day scaling relation between the stellar mass and central black hole mass of galaxies. The gaseous properties of galaxies and their haloes were not considered during the calibration.

\begin{table}
\centering
\begin{tabular}{l|rrrrr}
\hline
Identifier & $L$ & $N$ & $m_{\rm g}$ &  $\mathrm{\epsilon_{com}}$ & $\mathrm{\epsilon_{phys}}$\\ 
 & (cMpc)& & ($\Msun$)& (ckpc) & (pkpc) \\\hline
L025N0376 & 25 & $376^{3}$ & $1.81\times 10^6$ & 2.66 & 0.70 \\
L025N0752 & 25 & $752^{3}$ & $2.26\times 10^5$ & 1.33 & 0.35 \\
L100N1504 & 100 & $1504^{3}$ & $1.81\times 10^6$ & 2.66 & 0.70 \\ 
\hline
\end{tabular}
\caption{The box sizes and resolution details of the EAGLE simulations used in this study. The columns are: comoving box side length, $L$; number of DM particles (there is initially an equal number of baryon particles); the initial baryon particle mass; the Plummer-equivalent gravitational softening length in comoving units; the maximum proper softening length.}
\label{tab:sims}
\end{table}

EAGLE adopts values of the cosmological parameters derived from the initial Planck data release \citep{planck}, namely $\Omega_{0} = 0.307$,  $\Omega_{\mathrm{b}} = 0.04825$,  $\Omega_{\Lambda} = 0.693$, $\sigma_{\mathrm{8}} = 0.8288$, $n_{\mathrm{s}} = 0.9611$, $h = 0.6777$, $Y = 0.248$. Our analyses focus primarily on the EAGLE simulation of the largest cosmic volume, Ref-L100N1504, which follows a cubic periodic volume of side $L=100\cMpc$, realised with $N=1504^3$ collision-less DM particles of mass $m_{\mathrm{DM}} = 9.7 \times 10^{6} \mathrm{M_{\odot}}$, and an initially equal number of baryonic particles of mass $m_{\mathrm{g}} = 1.81 \times 10^{6} \mathrm{M_{\odot}}$. The Plummer-equivalent gravitational softening length is 1/25 of the mean interparticle separation ($\epsilon_{\rm com}=2.66\ckpc$), limited to a maximum proper length of $\epsilon_{\rm com}=0.7\pkpc$. We explore the numerical convergence of the morphology and orientation of the star-forming gas component of galaxies in Appendix \ref{sec:num_conv}, using the pair of high-resolution L025N0752 EAGLE simulations introduced by \citet{schaye15}. These follow a cosmic volume of $L=25\cMpc$ realised with $N=752^3$ particles of each species, with masses $m_{\mathrm{DM}} = 1.21 \times 10^{6} \mathrm{M_{\odot}}$ and $m_{\mathrm{g}} = 2.26 \times 10^{5}  \mathrm{M_{\odot}}$. For these simulations the Plummer-equivalent gravitational softening length is $\epsilon_{\rm com}=1.33\ckpc$, limited to a maximum proper length of $\epsilon_{\rm com}=0.35\pkpc$. The first of these simulations, Ref-L025N0752, uses the same `Reference' subgrid model parameters as the Ref-L100N1504, whilst the second, Recal-L025N0752, uses a model whose parameters were recalibrated to achieve a better match to the calibration diagnostics. A summary of the simulations used in this paper are given in Table~\ref{tab:sims}. 

The standard-resolution simulations marginally resolve the Jeans scales at the density threshold for star formation in the warm and diffuse photoionised ISM. They hence lack the resolution to model the cold, dense phase of the ISM explicitly, and so impose a temperature floor to inhibit the unphysical fragmentation of star-forming gas. This floor takes the form $T_{\mathrm{eos}}(\rho)$, corresponding to the equation of state $P_{\mathrm{eos}} \propto \rho^{4/3}_{\mathrm{g}}$ normalised to $T_{\mathrm{eos}} = 8 \times 10^{3} \mathrm{K}$ at $n_{\mathrm{H}} = 10^{-1}\mathrm{cm}^{-3}$. The temperature of star-forming gas thus reflects the effective pressure of the ISM, rather than its actual temperature. A drawback of the use of this floor is the suppression of the formation gas discs with scale heights much less than Jeans length of the gas on the temperature floor ($\sim 1\pkpc$). In Appendix~\ref{sec:eos_conv}, we explore the sensitivity of the star-forming gas morphology to the slope of the ISM equation of state, and the normalisation of the star formation law.

In a recent study, \citet{ludlow} demonstrated that the scale height of discs can be artificially increased by 2-body scattering of particles with unequal mass, as is the case here since we use (initially) equal numbers of baryon and DM particles, meaning that $m_{\rm dm}/m_{\rm b} \equiv (\Omega_0 - \Omega_{\rm b})/\Omega_{\rm b} \simeq 5.4$. The vertical support of the disc may also have physical causes, such as turbulence stemming from gas accretion and energy injection from feedback \citep{benitez18}, although it is likely that the these influences are artificially strong in the simulations. Therefore we caution that both the gas and stellar discs of galaxies in EAGLE are generally thicker than their counterparts in nature \citep[see also][]{trayford17}. We note however that these effects are unlikely to influence significantly the mutual alignment of the stellar and gaseous discs, nor their alignment with their parent DM halo. 

\subsection{Identifying and characterising haloes, subhaloes and galaxies}
\label{sec:halo_def}

We define galaxies as the cold baryonic component of gravitationally self-bound structures, identified by the application of the \textsc{subfind} algorithm \citep{springel01,dolag09} to DM haloes first identified with the friends-of-friends (FoF) algorithm (with a linking length of 0.2 times the mean interparticle separation). Subhaloes are identified as overdense regions in the FoF halo bounded by saddle points in the density distribution. Within a given FoF halo, the subhalo comprising the particle (of any type) with the lowest gravitational potential energy is defined as the central subhalo, others are then satellites. 

The position of galaxies is defined as the location of the particle in their subhalo with the lowest gravitational potential energy. The position of the central galaxy is used as a centre about which to compute the spherical overdensity mass \citep[see][]{lacey93}, $M_{200}$, for the adopted enclosed density contrast of 200 times the critical density, $\rho_{\rm c}$. In general, the properties of galaxies are computed by aggregating the properties of the appropriate particles located within $30\pkpc$ of the galaxy centre, as this yields stellar masses comparable to those recovered within a projected circular aperture of the Petrosian radius \citep[see][]{schaye15}. 

\subsection{Characterising the morphology and orientation of galaxy components}
\label{sec:redit}

Following \citet{thob19}, we obtain quantitative descriptions of the morphology of galaxies and their subhaloes by modelling the spatial distribution of their constituent particles as ellipsoids, characterised by their sphericity\footnote{\citet{thob19} used the flattening, $\epsilon = 1 - S$, rather than the sphericity but, as is clear from their definitions, the two are interchangeable.}, $S=c/a$, and triaxiality, $T=(a^2-b^2)/(a^2-c^2)$, parameters, where $a$, $b$ and $c$ are, respectively, the moduli of the major, intermediate and minor axes of the ellipsoid\footnote{\citet{thob19} present publicly available Python routines for this procedure at https://github.com/athob/morphokinematics.}. Therefore $S=0$ corresponds to a perfectly flattened (but potentially elongated) disc, and $S=1$ corresponds to a perfect sphere, whilst low and high values of $T$ correspond, respectively, to oblate and prolate ellipsoids. 

Axis lengths are given by the square root of the eigenvalues of a matrix describing the 3D mass distribution of the particles in question. The simplest choice is the mass distribution tensor \citep[e.g.][]{davisB, cole}, defined as:
\begin{equation} 
M_{ij}= \frac{\sum_{p} m_p r_{p,i} r_{p,j}}{\sum_{p} m_p},
\label{eq:simpinertiatensor}
\end{equation}
where the sum runs over all particles, $p$, comprising the structure, $r_{p,i}$ denotes the $i^{\rm th}$ component ($i,j=0,1,2$) of each particle's coordinate vector with respect to the galaxy centre, and $m_{\rm p}$ is the particle's mass. As has been widely noted elsewhere, the mass distribution tensor is often referred to as the moment of inertia tensor, as the two share common eigenvectors. 

There are several well-motivated alternative choices to the mass distribution tensor and, as per \citet{thob19}, we elect here to use an iterative form of the reduced inertia tensor \citep[see also][]{dubinski_carlberg, bett12, schneider}. The reduced form is advantageous because its suppresses a potentially strong influence on the tensor of structural features in the outskirts of galaxies, by down-weighting the contribution of particles at a large (ellipsoidal) radius. The use of an iterative scheme is further advantageous because it enables the scheme to adapt to particle distributions that deviate significantly from the initial particle selection. Since the latter is usually (quasi-)spherical, this is particularly relevant for strongly flattened or triaxial systems. This form of the tensor is thus:

\begin{equation}
\label{eq:redinertiatensor}
M^R_{ij}= \frac{\sum_{p} \frac{m_p}{\tilde{r}_p^2} r_{p,i} r_{p,j}}{\sum_{p} \frac{m_p}{\tilde{r}_p^2}},
\end{equation}
where $\tilde{r}_p$ is the ellipsoidal radius, and the superscript $R$ denotes that this is the reduced form of the tensor. In the first iteration, all particles of the relevant species within a spherical aperture of a prescribed radius, $r_{\mathrm{sph}}$, are considered. This yields a initial estimate of the axis lengths ($a,b,c$). In the next iteration, particles satisfying the following condition relating to the ellipsoidal distance are considered: 
\begin{equation}
     \widetilde{r}_{p}^{2} \equiv \frac{r_{p,a}^{2}}{\widetilde{a}^{2}} + \frac{r_{p,b}^{2}}{\widetilde{b}^{2}} + \frac{r_{p,c}^{2}}{\widetilde{c}^{2}} \leq 1,
\end{equation}
where $r_{p,a}$, $r_{p,b}$ and $r_{p,c}$ are the particle radii projected along the eigenvectors of the previous iteration, $\widetilde{a}, \widetilde{b}$ and $\widetilde{c}$ are the re-scaled axis lengths calculated as $\widetilde{a} = a \times r_{\mathrm{sph}}/(abc)^{1/3}$. This ensures the ellipsoid maintains a constant volume; in this respect, we differ from the scheme used by \citet{thob19}, who maintained a constant major axis length between iterations. We opt for this scheme to avoid artificial suppression of the major axis in cases of highly flattened geometry, which is more common when examining star-forming gas than is the case for stellar distributions. We note that our definition of $\tilde{r}_{p}^{2}$ differs with respect to that of \citet{thob19} by a factor $\tilde{a}^{2}$, and that often the normalisation factor $\sum_{p}m_{p}/\tilde{r}_{p}^{2}$ is not explicitly adopted in the definition of this tensor \citep[see e.g.][their equation 6]{bett12}. The axis lengths (and by extension, the shape parameters) recovered from the use of either form of the tensor are identical.

Iterations continue until the fractional change in the axis ratios $c/a$ and $b/a$ falls below 1 percent. If this criterion is not satisfied after 100 iterations, or if the number of particles enclosed by the ellipsoid falls below 10, the algorithm is deemed to have failed and the object's morphology is declared unclassified. We find a failure to converge only in cases of low particle number (e.g. subhaloes with very few gas or star particles) and, crucially, our selection criteria (Section \ref{sec:sample}) ensure that no subhaloes with unclassified morphologies are included in our sample. 

For consistency with the aperture generally used when computing galaxy properties by aggregating particle properties \citep[see e.g. Section 5.1.1. of][]{schaye15}, we adopt a radius of $r=30\pkpc$ for the initial spherical aperture. We use this aperture for all three matter types, star-forming gas, stars and DM, and note that for the latter, this focuses our morphology measurements towards halo centres, since haloes are in general much more extended than their cold baryons (see Section \ref{sec:mass_dist}). We retain the use of this aperture for the DM component in order to focus on the DM structure local to star-forming gas discs, and note that the \textit{global} morphology of DM haloes in EAGLE was presented by \citet{vel15a}. In Section \ref{sec:results_2d} we examine the 2D projected morphology and alignment of galaxies. When performing these measurements for star-forming gas and stars, we use an initial circular aperture of $r=\max(30\pkpc,2r_{1/2,\rm SF})$, where $r_{1/2,\rm SF}$ is the half-mass radius of star-forming gas bound to the subhalo. This ensures a robust morphological characterisation of the image projected by the most extended gas discs when viewed close to a face-on orientation.

Equation~\ref{eq:redinertiatensor} can be generalised to be weighted by any particle variable, rather than its mass. To crudely mimic the morphology of continuum-luminous regions, when computing the tensor for star-forming gas, we weight by their star formation rate (SFR) rather than their mass, since it is well-established that the relationship between SFR and radio continuum luminosity is broadly linear \citep[see e.g.][]{condon92, schober}. We do not consider radio continuum emission due to AGN, since this is not extended. \citet{pillepich19} recently employed a similar approach to assess the morphology and alignment of H$\alpha$-luminous regions of star-forming galaxies in the TNG50 simulation, via the use of the SFR as a proxy for the H$\alpha$ luminosity. The recovered shape parameters and orientation are little changed with respect to the use of particle mass as the weighting variable, or indeed a uniform weighting, largely because the SFR of particles scales as $\dot{m}_\star \propto P^{1/5}$ for a Kennicutt-Schmidt law with index $n_{\rm s} = 1.4$ \citep[see][]{schaye_DV} and the pressure distribution of star-forming gas particles is relatively narrow: at $z=0$, the 10th and 90th percentiles of the pressure of star-forming particles in the Ref-L100N1504 volume spans less than two decades in dynamic range.

We define the orientation of galaxies and subhaloes as the unit vector parallel to the minor axis of the best-fitting ellipsoid, and hence measure the relative alignment of structures as the angle between these unit vectors. We note that it is more typical in the literature to use the unit vector parallel to the major axis; this is arguably the best-motivated choice for describing the alignment of systems that are in general prolate (e.g. DM haloes), since in such systems the major axis is the most `distinct'. In contrast, it is the minor axis that is the most distinct in systems that are preferentially oblate, as is the case for a flattened disc. In Section \ref{sec:kin_morph_align} we examine the correspondence between the morphological and kinematic axes of the star-forming gas distribution; we define the latter as the unit vector parallel to the angular momentum vector of all star-forming gas particles located within $30\pkpc$ of the galaxy centre.  

\subsection{Sample selection}
\label{sec:sample}

We identify subhaloes comprising a minimum of 100 each of star-forming gas particles, stellar particles and DM particles. This numerical threshold is motivated by tests, presented in Appendix \ref{sec:min_part}, that assess the fractional error on shape parameters induced when performing the measurement on subsamples, randomly selected and of decreasing size, of the particles comprising exemplar subhaloes. These tests indicate that a minimum of 100 particles are needed to recover a measurement error of the flattening of star-forming gas discs of less than $10$ percent, when using the iterative reduced inertia tensor. As noted by \citet{thob19}, the sphericity and triaxiality shape parameters are poor descriptors of systems that deviate strongly from axisymmetry, so we excise subhaloes with strongly non-axisymmetric star-forming gas distributions.  We quantify this characteristic by adapting the method of \citet{trayford19}, binning the mass of star-forming gas into pixels of solid angle about the galaxy centre using \textsc{Healpix} \citep{gorski05}. The asymmetry of the star-forming gas distribution, $A_{\rm 3D}$, is then computed by summing the (absolute) mass difference between diametrically opposed pixels and normalising by the total star-forming gas mass. As per \citet{trayford19}, we use coarse maps of 12 pixels, and exclude systems with $A_{\rm 3D}^{\rm SFG} > 0.6$. This criterion excises 534 subhaloes, mostly of low mass, and leaves us with a sample of 6,764 subhaloes at $z=0$. 

Our selection criteria, in particular the requirement for subhaloes to be comprised of at least 100 particles each of stars and star-forming gas, impose a strong selection bias at low halo masses. In practice, for simulations at the resolution of the EAGLE Ref-L100N1504 simulation, the criteria dictate that subhaloes host a galaxy with a minimum stellar of mass of $\sim 10^8\Msun$ and a minimum SFR of $\simeq 6\times 10^{-2} \mathrm{M_{\odot}yr^{-1}}$, where the latter assumes the star-forming particles have a density of $0.1\cmcubed$ and pressure corresponding to a temperature of $8000\K$. This corresponds to a specific star formation rate (sSFR) of $6\times10^{-10}\peryr$ for the lowest (stellar) mass galaxies, a value that is above the canonical threshold separating the blue cloud of star-forming galaxies and the red sequence of quenched counterparts \citep[e.g.][]{schawinski2014}. Our selection criteria result in the selection of approximately (0.1, 10, 80) percent of all subhaloes of mass $\log_{10} (M_{\rm sub}/\Msun) \sim (10,11,12)$, respectively, corresponding to approximately (16, 65, 60) percent of all subhaloes of stellar mass $\log_{10} (M_{\rm \star}/\Msun) \sim (9,10,11)$.

\subsection{Mass distribution profiles}
\label{sec:mass_dist}

\begin{figure}
\centering
\hspace{-0.2cm}
     \includegraphics[width = 0.45\textwidth]{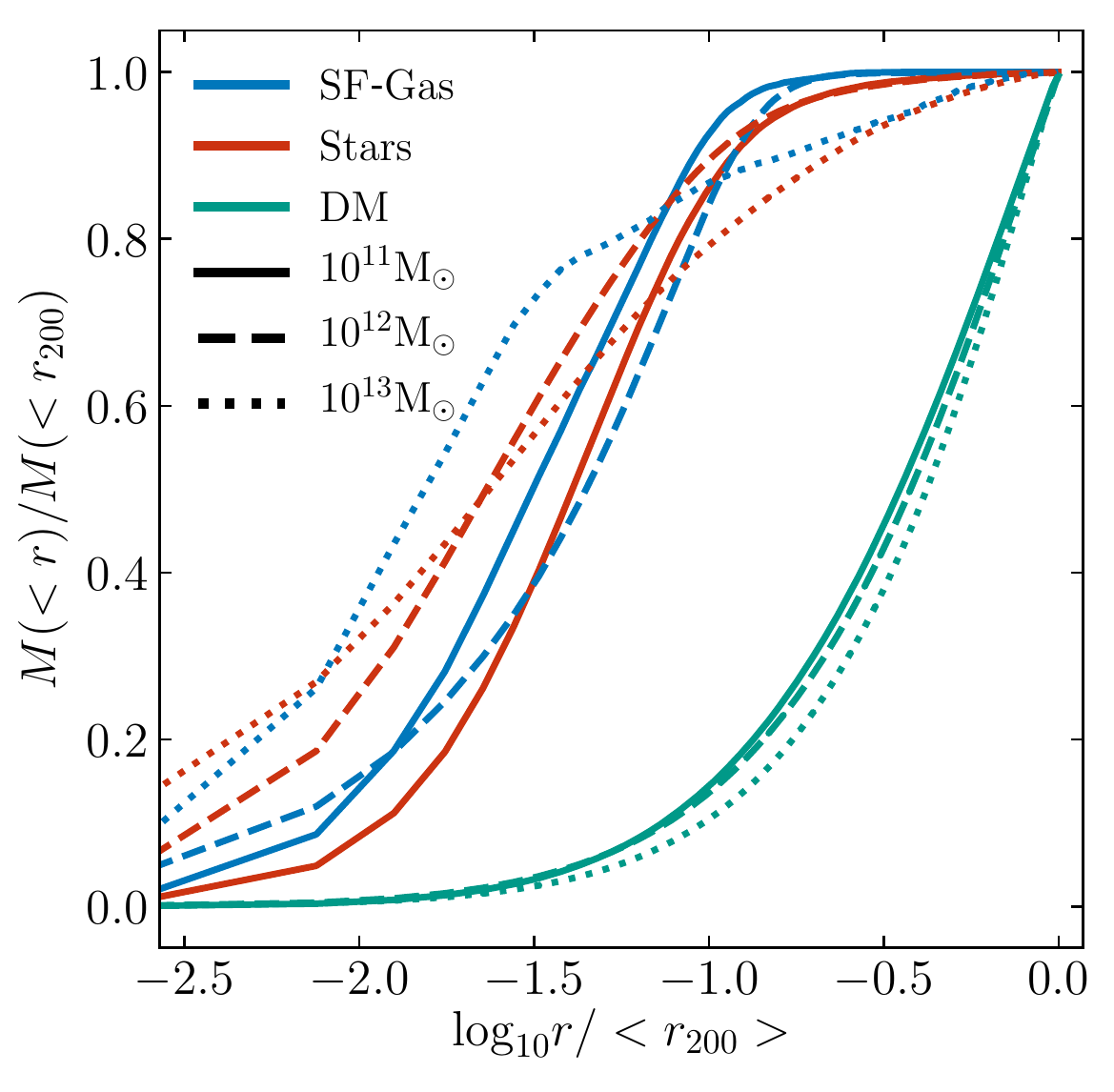}
\caption{Mean, spherically averaged cumulative radial mass distribution profiles of the star-forming gas (blue curve), stars (red), and DM (green) of central subhaloes within our sample that have a halo mass $M_{200} \sim 10^{11}\mathrm{M_{\odot}}$ (solid curve), $10^{12}\mathrm{M_{\odot}}$ (dashed), and $10^{13}\mathrm{M_{\odot}}$ (dotted). The distributions are normalised relative to the total mass of each component within $r_{200}$. Star-forming gas is much more centrally concentrated than dark matter at all masses.}
\label{fig:mass_dist}
\end{figure}

Prior studies have demonstrated that the shape and orientation of stars and DM in haloes can vary significantly as a function of radius \citep[see e.g.][]{vel15a}. Fig. \ref{fig:mass_dist} shows the mean, spherically averaged, cumulative radial mass distribution profiles of the star-forming gas (blue curve), stars (red), and DM (green) comprising present-day central subhaloes with halo mass in ranges $M_{200} \sim10^{11}\mathrm{M_{\odot}}$ (solid curves), $10^{12}\mathrm{M_{\odot}}$ (dashed), and $10^{13}\mathrm{M_{\odot}}$ (dotted). As might be na\"ively expected, the baryonic components are much more centrally concentrated than the DM, in each of the subhalo mass bins: the median half-mass radius of star-forming gas is $(3,4.5,1.5)$ percent of $r_{200}$ for the low, middle and high mass bins respectively, compared with $ (35,37,42)$ percent of $r_{200}$ for the DM\footnote{The figures for the low subhalo mass bin are significantly influenced by our sample selection criteria: removal of the minimum particle number criterion results in the inclusion of systems with less-extended star-forming gas distributions, and further reduces the characteristic half-mass radius of the star-forming gas.}. Owing to this central concentration of the star-forming gas, we do not consider here how the shape parameters of the star-forming gas distribution change in response to the use of an initial aperture that envelops an ever-greater fraction of the virial radius. 

\section{The morphology of star-forming gas}
\label{sec:results_morphology}

\begin{figure*}
\centering
\hspace{-0.2cm}
     \includegraphics[width = 0.95\textwidth]{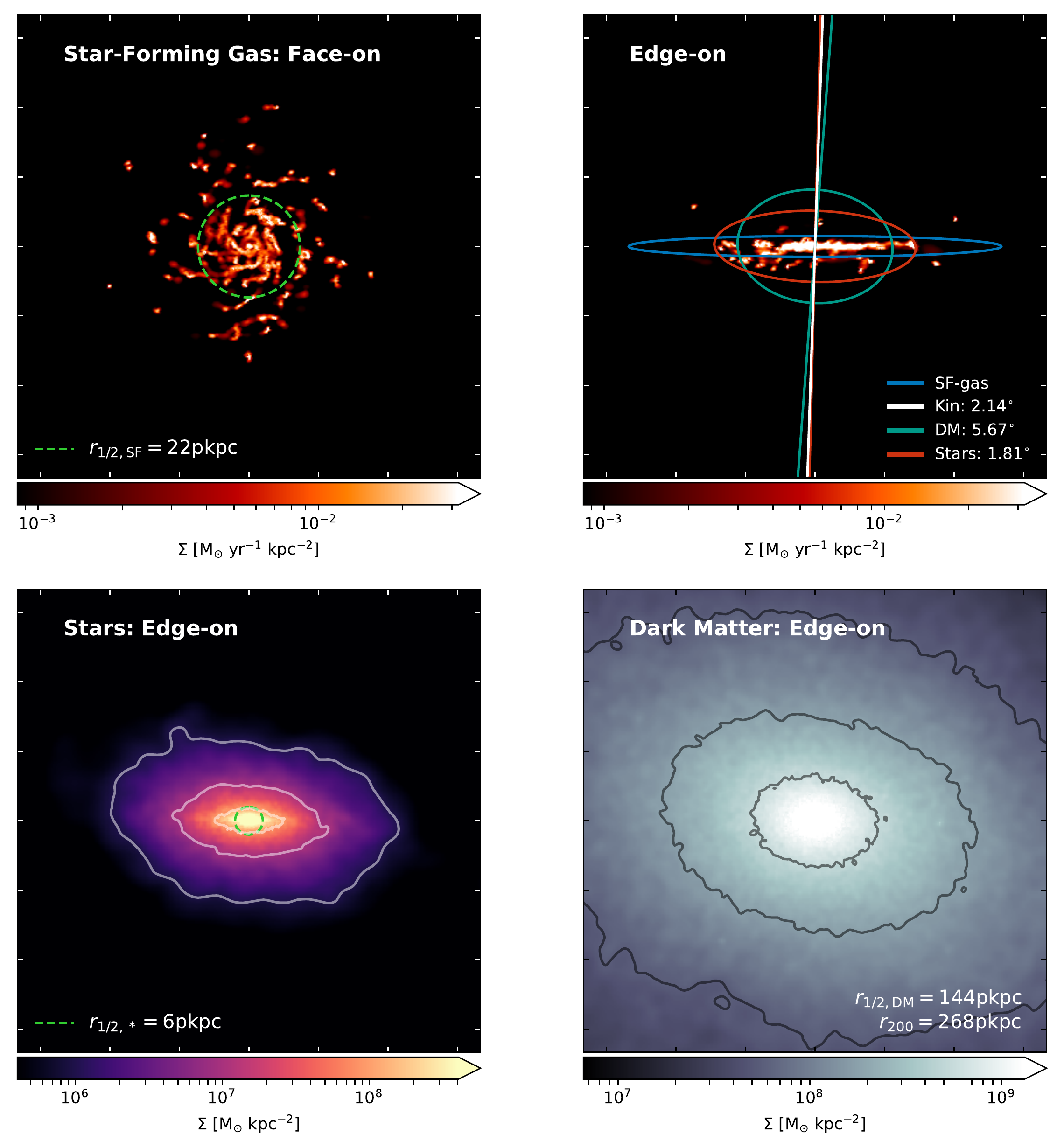}
     \caption{The star-forming gas, stars and dark matter (DM) comprising a star-forming central galaxy drawn from Recal-L025N0752, with stellar mass $M_{\star} = 10^{10.5}\Msun$. Each panel is $200\pkpc$ on a side. The galaxy's subhalo mass is $M_{\mathrm{sub}}=10^{12.4}\Msun$, and its sSFR is $\dot{M_\star}/M_\star= 10^{-10.5}\mathrm{yr^{-1}}$. The upper panels show the star formation rate surface density, a simple proxy for the radio continuum surface brightness, viewed face-on and edge-on. The green circle in the upper left-hand panel denotes the spherical half-mass radius of star-forming gas within $30\pkpc$. Ellipsoids in the upper right-hand panel show projections of the best-fitting ellipsoids of the three matter components recovered by the iterative reduced inertia tensor. Overlaid solid lines show the minor axis of the stars (red), and DM (green), whilst the white line corresponds to the rotation axis. The SF-gas is much flatter ($S = 0.06$) than the stars ($S = 0.35$) and DM ($S = 0.73$). The lower panels show the surface densities of stars (left-hand panel) and DM (right-hand panel). Overlaid contours denote surface densities of $\mathrm{log_{10}}(\Sigma_{\star}) = 6$, $7$, $8$ $\mathrm{\Msun kpc^{-2}}$ and $\mathrm{log_{10}}(\Sigma_{\mathrm{DM}}) = 7.75, 8.25, 8.75$ $\mathrm{\Msun kpc^{-2}}$ for the stars and DM respectively. The spherical half-mass radii for the stars and dark matter are $6\pkpc$ and $144\pkpc$ respectively. These images have been made using the publicly available code \textsc{Py-SPHViewer} \citep{benitez}}
\label{fig:four_pan}
\end{figure*}

We begin with an examination of the morphology of star-forming gas associated with subhaloes. To illustrate visually how the method described in Section \ref{sec:redit} yields shape and orientation diagnostics for the simulated galaxies, we show in Fig. \ref{fig:four_pan} the star formation rate surface density, $\Sigma_{\rm SFR}$, of star-forming gas (upper row), in face-on and edge-on views, and the mass surface density of stars ($\Sigma{\star}$, bottom left-hand panel) and DM ($\Sigma_{\rm DM}$, bottom right-hand panel) of a present-day star-forming galaxy from Recal-L025N0752. The galaxy is taken from the high-resolution Recal-L025N0752 run, and its stellar mass is $M_\star = 10^{10.5}\Msun$, with a subhalo mass of $M_{\mathrm{sub}}=10^{12.4}\Msun$. The galaxy's sSFR is $\dot{M_\star}/M_\star = 10^{-10.5}\mathrm{yr^{-1}}$, and it exhibits reasonably strong rotational support: stars residing within $30\pkpc$ of its centre of potential have a significant fraction of their kinetic energy invested in corotation ($\kappa_{\rm co}^{\star}=0.44$). For reference, \citet{correa17} argue that $\kappa_{\rm co}^{\star}>0.4$ is a useful and simple criterion for identifying star-forming disc galaxies in EAGLE. The star-forming gas exhibits a very high degree of rotation support, $\kappa_{\rm co}^{\rm SF}=0.97$.

The field of view of each panel is $200\pkpc$, and overlaid dashed green circles denote the half-mass radius of the matter type in question. Edge-on images are aligned such that horizontal and vertical image axes are parallel to the major and minor axes, respectively, of the star-forming gas distribution. In the upper right-hand panel, coloured ellipses correspond to projections of the best-fitting ellipsoids describing the respective matter components, whilst the solid coloured lines show the (projected) minor axes of the stellar and DM distributions, and the white line shows the projected rotation axis of the star-forming gas. Contours overlaid on the stellar and DM surface density images correspond to surface densities of $\mathrm{log_{10}}(\Sigma_{\star}) = 6$, $7$, $8$ $\mathrm{\Msun kpc^{-2}}$ and $\mathrm{log_{10}}(\Sigma_{\mathrm{DM}}) = 7.75, 8.25, 8.75$ $\mathrm{\Msun kpc^{-2}}$, respectively. 

As expected for a galaxy whose gas disc has strong rotational support, the star-forming gas distribution is much more flattened than the corresponding distributions of stars and DM. In this example, the distributions of the three matter components are well aligned: the minor axis of the star-forming gas is misaligned with respect to that of the stars by $\simeq 2$ deg and the DM by $\simeq 6$ deg. As shown in Appendix \ref{sec:min_part}, these offsets are comparable to the measurement uncertainty for well-resolved and well-sampled structures. The rotational axis of the star-forming gas is also closely aligned with the minor axis in this example, as na\"ively expected for an extended, rotationally supported disc.

\subsection{Shape parameters as a function of subhalo mass}
\label{sec:shape_by_mass}

\begin{figure*}
\centering
     \includegraphics[width = 0.9\textwidth]{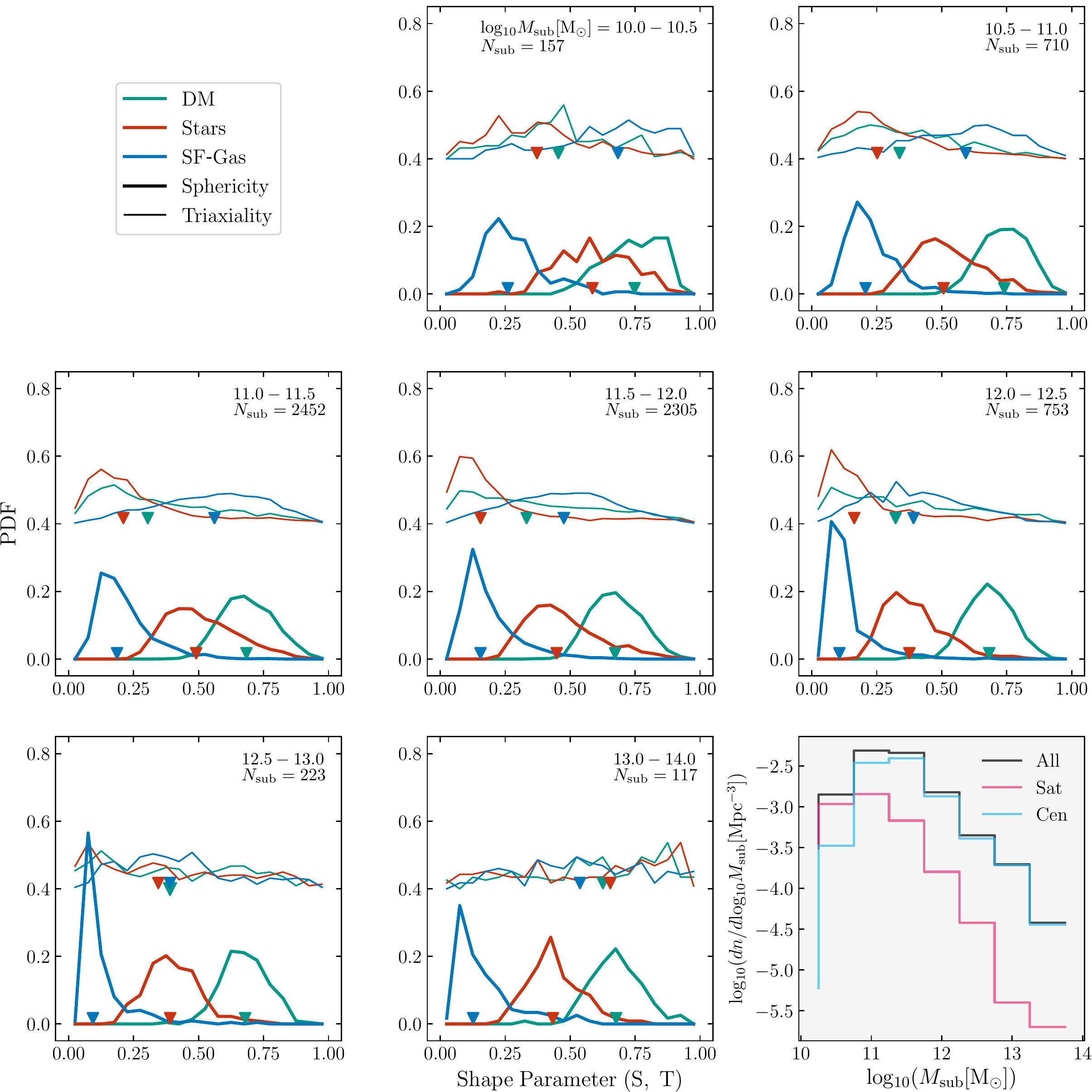}
\caption{Probability distribution functions (PDFs) for the sphericity and triaxiality of the star-forming gas (blue), stars (red), and DM (green) comprising sampled subhaloes in Ref-L100N1504 at $z = 0$. Subhaloes are binned by their total mass. The triaxiality PDFs have been raised artificially by $0.4$ for clarity. The arrows represent the location of the median for each distribution. The bottom right-hand panel shows volume density function of the satellite and central galaxies. For all subhalo masses, the star-forming gas is significantly more flattened than the stars and DM, and the distribution of the sphericity parameter for star-forming gas is particularly narrow in $\sim L^\star$ galaxies. Star-forming gas does not exhibit a characteristic triaxiality, spanning a wide range of $T$ at all subhalo masses.}
\label{fig:pdfs}
\end{figure*}

\begin{figure}
\centering
\hspace{-0.2cm}
         \includegraphics[width = 0.45\textwidth]{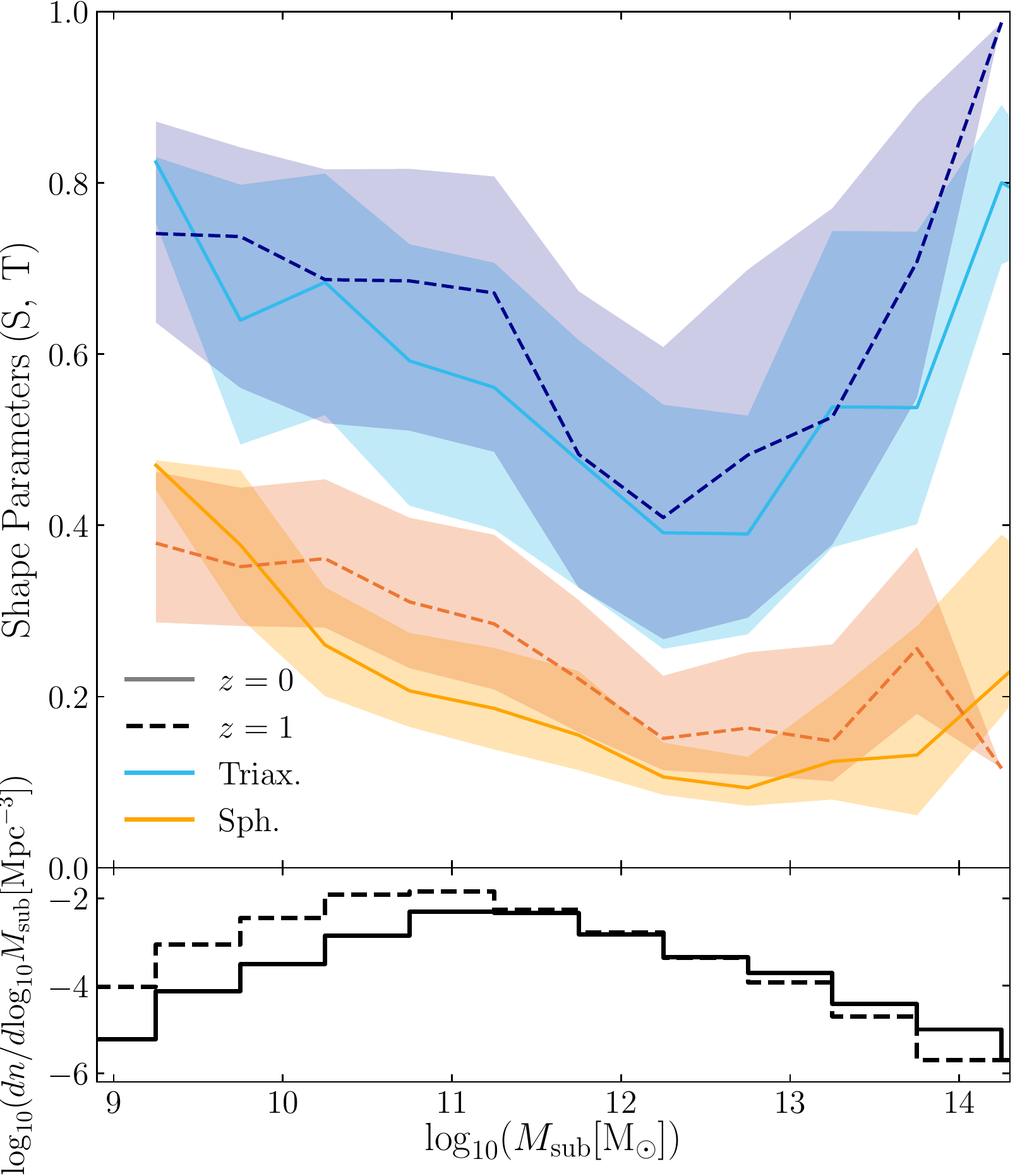}
\caption{The median sphericity and triaxiality of the star-forming gas distribution of subhaloes, as a function of subhalo mass, at $z = 0$ and $z=1$ in bins of 0.5 dex. Shaded regions denote the interquartile range. The orange and blue curves correspond to the sphericity and triaxiality, whilst solid and dashed curves correspond to $z = 0$ and $z=1$, respectively. The lower panel shows the volumetric mass function of the sampled subhaloes.}
\label{fig:S_and_T_vs_mass}
\end{figure}

Fig. \ref{fig:pdfs} shows probability distribution functions (PDFs) of the shape parameters of the star-forming gas (blue curves), stellar (red) and DM (green) distributions of the subhaloes comprising our sample from Ref-L100N1504 at $z=0$. We reiterate that measurements of the stellar and DM distributions are included here, despite being previously presented for EAGLE subhaloes by \citet{vel15a}, because we use an alternative form of the mass distribution tensor. Thick and thin lines represent the sphericity and triaxiality parameters respectively. Each panel shows subhaloes split by total mass in bins of 0.5 dex, spanning $M_{\rm sub} = 10^{10} - 10^{14}\Msun$. For clarity, the PDFs of triaxiality have been artificially elevated in the vertical axis by an increment of 0.4. Down arrows denote the median value of each distribution. The bottom right-hand panel shows the volumetric subhalo mass function, split into central and satellite subhalo populations, highlighting that the sample is dominated by central galaxies at all subhalo masses except for the lowest mass bin. For clarity, we also show the median values of the shape parameters for star-forming gas as a function of subhalo mass in Fig. \ref{fig:S_and_T_vs_mass}. The solid and dashed curves of that plot correspond to the samples, identified as discussed in Section \ref{sec:sample}, at $z=0$ and $z=1$, respectively. The lower panel of the figure shows the subhalo volumetric mass function at the two epochs.

These figures show that the distribution of sphericities of star-forming gas distributions is peaked at relatively low values for all subhalo masses, but with a long tail towards high $S$ (i.e. quasi-spherical systems). The median value of the distributions, which is qualitatively similar to the peak value of the distribution, declines from $\tilde{S}\simeq 0.25$ for subhaloes of $M_{\rm sub}\sim 10^{10}\Msun$, to a minimum of $\tilde{S}\simeq 0.1$ at $M_{\rm sub}\sim 10^{12.5}\Msun$. The sphericity of the star-forming gas is therefore systematically lower than is the case for that of the stars, and much more so than is the case for the DM, consistent with the na\"ive expectation that this dissipational component is found primarily in flattened discs. Broadly, the peaks of the sphericity PDFs of stars and DM are found at $S\simeq 0.3-0.5$ and $S\simeq 0.7-0.75$, respectively, irrespective of subhalo mass. \citet{thob19} noted that present-day galaxies whose stellar component exhibit a sphericity of $S \lesssim 0.6$ generally exhibit stellar corotation kinetic energy fractions of $\kappa_{\rm co}^{\star} > 0.4$ and so correspond broadly to blue, star-forming disc galaxies \citep{correa17}. Despite our use of an initial $30\pkpc$ aperture for the mass tensor, the median values of the sphericity of the stars and DM are broadly consistent with those recovered by \citet{vel15a} when applying the standard mass distribution tensor to the entirety of EAGLE subhaloes, and those recovered by \citet{tenneti14} for subhaloes in the MassiveBlack-II simulation in the mass range for which our respective selection criteria recover broadly similar samples of galaxies ($M_{\rm sub} \gtrsim 10^{11}\Msun$). Similarly, the distribution of sphericities of star-forming gas discs are consistent with those recovered by \citet{pillepich19} when applying the standard mass distribution tensor to galaxies in the TNG50 simulation. We remark that we have also computed the morphology of star-forming gas structures using an iterative form of the simple mass tensor (equation~\ref{eq:simpinertiatensor}), and do not find a significant systematic change.

\begin{table}
\centering
\begin{tabular}{l|l|l|l|l|l|l}
\hline
$\log_{10} M_{\rm sub}$ & \multicolumn{3}{c}{Sphericity, $S$} & \multicolumn{3}{c}{Triaxiality, $T$} \\
$[\Msun]$ & SF-gas & Stars & DM & SF-gas & Stars & DM  \\
\hline 
10.0-10.5 & 0.13 & 0.21 & 0.16 & 0.28 & 0.27 & 0.26\\
10.5-11.0 & 0.11 & 0.18 & 0.13 & 0.31 & 0.24 & 0.31 \\
11.0-11.5 & 0.12 & 0.19 & 0.14 & 0.31 & 0.23 & 0.36\\
11.5-12.0 & 0.12 & 0.18 & 0.13 & 0.29 & 0.2 & 0.41\\
12.0-12.5 & 0.06 & 0.15 & 0.12 & 0.29 & 0.23 & 0.4\\
12.5-13.0 & 0.06 & 0.14 & 0.11 & 0.26 & 0.48 & 0.48\\
13.0-14.0 & 0.14 & 0.13 & 0.13 & 0.37 & 0.51 & 0.38\\
\hline 
\end{tabular}
\caption{Interquartile ranges of the distributions of the sphericity ($S$) and triaxiality ($T$) shape parameters of the star-forming gas (SF-gas), stars and dark matter (DM) comprising subhaloes in our sample, as a function of subhalo mass.}
\label{tab:iqr}
\end{table}

The sphericity of star-forming gas is most uniform in subhaloes of intermediate mass, $M_{\rm sub} \sim 10^{11.5-13}\Msun$. In such structures, the distribution of $S$ is strongly peaked at low values corresponding to flattened discs, albeit with a long tail to more spherical configurations. Owing to this asymmetry, which is most prominent for the star-forming gas, we quantify the diversity of the shape parameter distributions via the interquartile range (IQR) rather than their variance (see Table \ref{tab:iqr}). The IQR of the star-forming gas sphericity decreases from $0.13$ for subhaloes of $\log_{10} (M_{\rm sub}/\Msun)=10-10.5$ to a minimum of $0.06$ for subhaloes of $\log_{10} (M_{\rm sub}/\Msun)=12-13$, before increasing again to $0.14$ for the most massive haloes in our sample. The greater diversity in low-mass subhaloes is driven largely by stochasticity in the structure of star-forming gas, with star formation in many low-mass galaxies being confined to a small number of gas clumps rather than being distributed throughout a well-defined disc. In massive subhaloes, cold gas discs are readily disturbed by outflows driven by efficient AGN feedback \citep[see e.g.][]{bower17,oppenheimer19}, and are less readily replenished with high-angular momentum gas from coherent circumgalactic inflows \citep[see e.g.][]{davies20, dcp20}. 

A potentially surprising finding highlighted by Figs. \ref{fig:pdfs} and \ref{fig:S_and_T_vs_mass} is that the characteristic morphology of present-day star-forming gas discs can deviate significantly from that of a disc. The characteristic triaxiality of star-forming gas in subhaloes of mass $M_{\rm sub}\sim 10^{12-12.5}\Msun$ is $T < 0.5$, consistent with a flattened, oblate spheroid. Subhaloes of all masses exhibit a broad distribution of $T$, in marked contrast with that of $S$, and for subhaloes in the lower and higher mass bins, the median value is $T > 0.5$, signifying that the characteristic morphology is prolate, such that even though the structures are flattened, their isodensity contours when viewed face-on deviate significantly from circular. A similar finding from the TNG50 simulation was recently reported by \citet{pillepich19}. Inspection of face-on projections of the star-forming gas surface density highlights that this behaviour again stems primarily from the stochasticity of star-forming gas structure in low-mass subhaloes. In more massive subhaloes, stochasticity is also relevant, owing to the efficient disruption of well-sampled cold gas discs by AGN feedback. However we note that the stellar and DM components tend towards more prolate configurations in more massive subhaloes \citep[as has been widely reported elsewhere, e.g.][]{tenneti14, vel15a}, suggesting that the morphology of the gravitational potential may influence that of the cold gas. We examine this further in Section \ref{sec:correlated_shape}.

As previously noted by \citet{vel15a}, the triaxiality of the stars and DM in EAGLE subhaloes increases as a function of the subhalo mass, such that these components in the most-massive structures are strongly prolate. We note that our quantitative measures are however slightly lower than those reported by \citet{vel15a}, owing to our use of an initial $30\pkpc$ aperture and the reduced inertia tensor, which ascribes less weight to morphology of these structures at large (elliptical) radius. It is well established from prior studies that the condensation of baryons in halo centres drives the morphology towards a more spherical configuration than is realised in dark matter-only simulations \citep[see e.g.][]{dubinski, katz94, kazantzidis, springel04, zemp}.

\subsection{Shape parameters at $z > 0$}

\begin{figure*}
\centering
     \includegraphics[width = 0.95\textwidth]{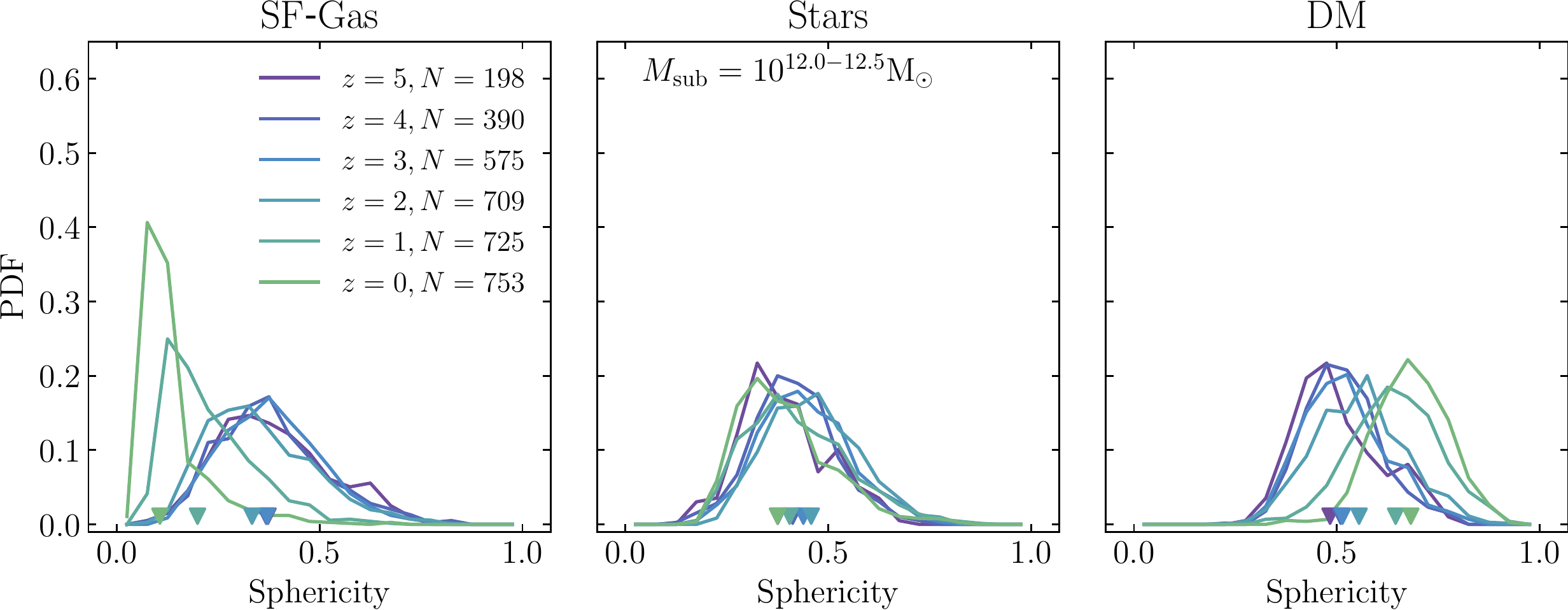}
     \caption{The sphericities of the three matter components (star-forming gas, stars and DM, from left to right, respectively) comprising subhaloes of present-day mass $\log_{10}(M_{\rm sub}/\Msun)=12-12.5$, and their main progenitor subhaloes at $z=(1,2,3,4,5)$. The latter are included only whilst still satisfying the $z=0$ selection criteria, so the sample size $N$, shown in the legend of the left-hand panel, is a monotonically declining function of redshift. The arrows represent the location of the median for each distribution. The characterisation of the matter distributions with the iterative reduced inertia tensor indicates that DM becomes more spherical at late cosmic epochs, the stellar distribution evolves mildly towards a more flattened configuration, and the star-forming gas typically evolves strongly towards a very flattened configuration by the present-day.}
\label{fig:prog}
\end{figure*}

We now turn to the morphology of subhaloes at $z>0$, for which we take two approaches. First, we identify subhaloes at $z=1$ (which, for our adopted cosmogony, corresponds to a lookback time of $8.1\Gyr$) that satisfy the selection criteria specified in Section \ref{sec:sample}, and compare the shape parameters of the samples at these epochs. We subsequently explore the evolution of the shape parameters of the main progenitors of subhaloes that satisfy the selection criteria at $z=0$. Clearly, these approaches require the examination of increasingly dissimilar subhalo samples as one advances to higher redshift.

The evolution of the characteristic morphology of the star-forming gas for identically selected samples at $z=0$ and $z=1$, respectively, can be assessed from comparison of the solid and dashed curves of Fig. \ref{fig:S_and_T_vs_mass}. These curves denote the median values of the shape parameters (sphericity in orange, triaxiality in blue) as a function of subhalo mass, whilst the shaded regions correspond to the interquartile range. The darker (lighter) shaded areas for each parameter correspond to $z = 1$ ($z=0$). 

It reveals that cold gas structures of fixed subhalo mass, for $\log_{10} (M_{\rm sub}/\Msun) \gtrsim 10$, are slightly more spherical (i.e. less flattened) at $z=1$ than the present-day. However this difference ($<0.1$ for all subhalo masses) is smaller than, or comparable to, the interquartile range of $S$ at either epoch, which varies between $0.03$ and $0.22$ at $z=0$ and $0.11$ and $0.19$ at $z=1$, over the subhalo mass range from $\log_{10} (M_{\rm sub}/\Msun)=9-14$. Similarly, the star-forming gas in subhaloes of the same mass tends to be less oblate / more prolate at $z=1$ than the present-day, but again the difference is small in comparison to the scatter at fixed subhalo mass. We note that the trend for sphericity is in marked contrast with the qualitative behaviour of the DM, for which there is a consensus that structures become more spherical with advancing cosmic time \citep[see e.g.][]{bryan13,tenneti14,vel15a}.

Fig. \ref{fig:prog} shows the sphericity PDFs of the three matter components (star-forming gas, stars and DM from left to right, respectively) of the main progenitor subhalo, at $z=(0,1,2,3,4,5)$, of present-day central subhaloes with mass $\log_{10} (M_{\rm sub}/\Msun)=12.0-12.5$. Such subhaloes broadly correspond to those that host present-day $\sim L^\star$ galaxies. The progenitors are identified using the \textsc{D-Trees} algorithm \citep{jiang}; a full description of its application to the EAGLE simulations is provided by \citet{qu}. The standard $30\pkpc$ aperture is used at all redshifts\footnote{We have assessed the impact of using an adaptive aperture of initial spherical radius $r=0.15 r_{200}(z)$, to account for the decreasing physical size of progenitors at early times, and do not recover significant differences.}. Progenitor subhaloes are included in the $z>0$ samples only while they still satisfy the selection criteria concerning particle number and asymmetry, to ensure that a robust measurement of their shape parameters can be made. As such, the sample size, $N$, is a monotonically declining function of redshift, as denoted in the legend of the left-hand panel of the figure.

We saw from Fig. \ref{fig:pdfs} that present-day galaxies hosted by subhaloes in this mass range typically exhibit strongly flattened ($S\simeq 0.1$) star-forming gas discs. The left-hand panel of Fig.~\ref{fig:prog} highlights that, although star-forming gas discs are predominantly flattened\footnote{For context, we reiterate that, as noted in Section \ref{sec:shape_by_mass}, present-day galaxies with a stellar component sphericity of $S \lesssim 0.6$ are broadly equivalent to star-forming disc galaxies.} even at early epochs, the median sphericity at $z=5$ is $\tilde{S}_{\rm SF-gas} \simeq 0.37$. The star-forming gas of the main progenitor becomes increasingly flattened with advancing cosmic time, but the emergence of strongly flattened discs ($S \lesssim 0.2$) is generally limited to $z<2$: the median sphericity evolves from $\tilde{S}_{\rm SF-gas} \simeq 0.33$ at $z=2$ to $\tilde{S}_{\rm SF-gas} \simeq 0.1$ at $z=0$. The strong evolution of the star-forming gas sphericity of these progenitors is broadly coincident with the growth of the gas disc's median scale length, which grows only from $\simeq 2\pkpc$ to $\simeq 3\pkpc$ from $z=5$ to $z=2$, but by $z=0$ reaches $\simeq 9\pkpc$. The decrease in the accretion rate (of all matter types) onto the galaxy+halo ecosystem at later epochs \citep[see e.g.][]{fakhouri10,vandevoort11} likely also results in a steady decline of the scale height of the gas disc \citep[e.g.][]{benitez18}, further contributing to decrease in sphericity. 

Strong evolution of the structural parameters of star-forming gas was similarly reported by \citet{pillepich19} based on analysis of the TNG50 simulation. Those authors noted that the evolution of the flattening (`disciness' in their terminology, since they also examined kinematic descriptions) of both the star-forming gas and stars increases over time, but that the evolution for the former is much more pronounced than the latter. The same behaviour is evident in EAGLE, as is clear from inspection of the centre panel of Fig.~\ref{fig:prog}, which shows that the sphericity of the stellar component of the progenitors of our present-day $\sim L^\star$ galaxy sample is largely insensitive to redshift. The majority of the galaxies comprising our sample remain actively star-forming at $z=0$, and are characterised by flattened discs ($\tilde{S}_{\star} \simeq 0.4-0.45$ at all redshifts examined). Such galaxies will therefore have assembled primarily via in-situ star formation \citep{qu} and will not have experienced the strong morphological evolution that typically follows internal quenching \citep[see e.g.][]{davies20,dcp20}. \citet{furlong17} showed that the half-mass radius of the stellar component of present-day star-forming $\sim L^\star$ galaxies grows only from $\simeq 10^{0.5}\pkpc$ to $\simeq 10^{0.85}\pkpc$ between $z=2$ and $z=0$. 

As noted above, it has been shown elsewhere that DM haloes, even in the absence of dissipative baryon physics, tend to become more spherical with advancing cosmic time \citep[e.g.][]{bryan13,tenneti14,vel15a}, in marked contrast to the behaviour seen for the star-forming gas. The right-hand panel of Fig. \ref{fig:prog} shows that this effect is clearly seen for the host subhaloes of present-day $\sim L^\star$ galaxies, even when focusing primarily on the halo centre by defining the shape parameters via the use of the iterative reduced mass tensor. The resolved progenitors exhibit a median sphericity of $\tilde{S}_{\rm DM} \simeq 0.5$ at $z=5$, and this median increases monotonically to $\tilde{S}_{\rm DM}\simeq 0.7$ at $z=0$. 

Besides the evolution of the median sphericity of the matter components, it is interesting to consider the evolution of their diversity. Since the PDFs can exhibit significant asymmetry, we characterise this diversity using the interquartile range. Whilst the IQR of the star-forming gas sphericity decreases markedly at later cosmic epochs (c.f $0.18$ at $z=5$ to $=0.06$ at $z=0$), that of the stars and the DM remain components remain largely unchanged from $z=5$ to $z=0$, with values of 0.13 to 0.14 for the stars and 0.14 to 0.12 for the DM.

\subsection{Correspondence of star-forming gas and stellar structure}
\label{sec:correlated_shape}

\begin{figure*}
\centering
     \includegraphics[width = 1.0\textwidth]{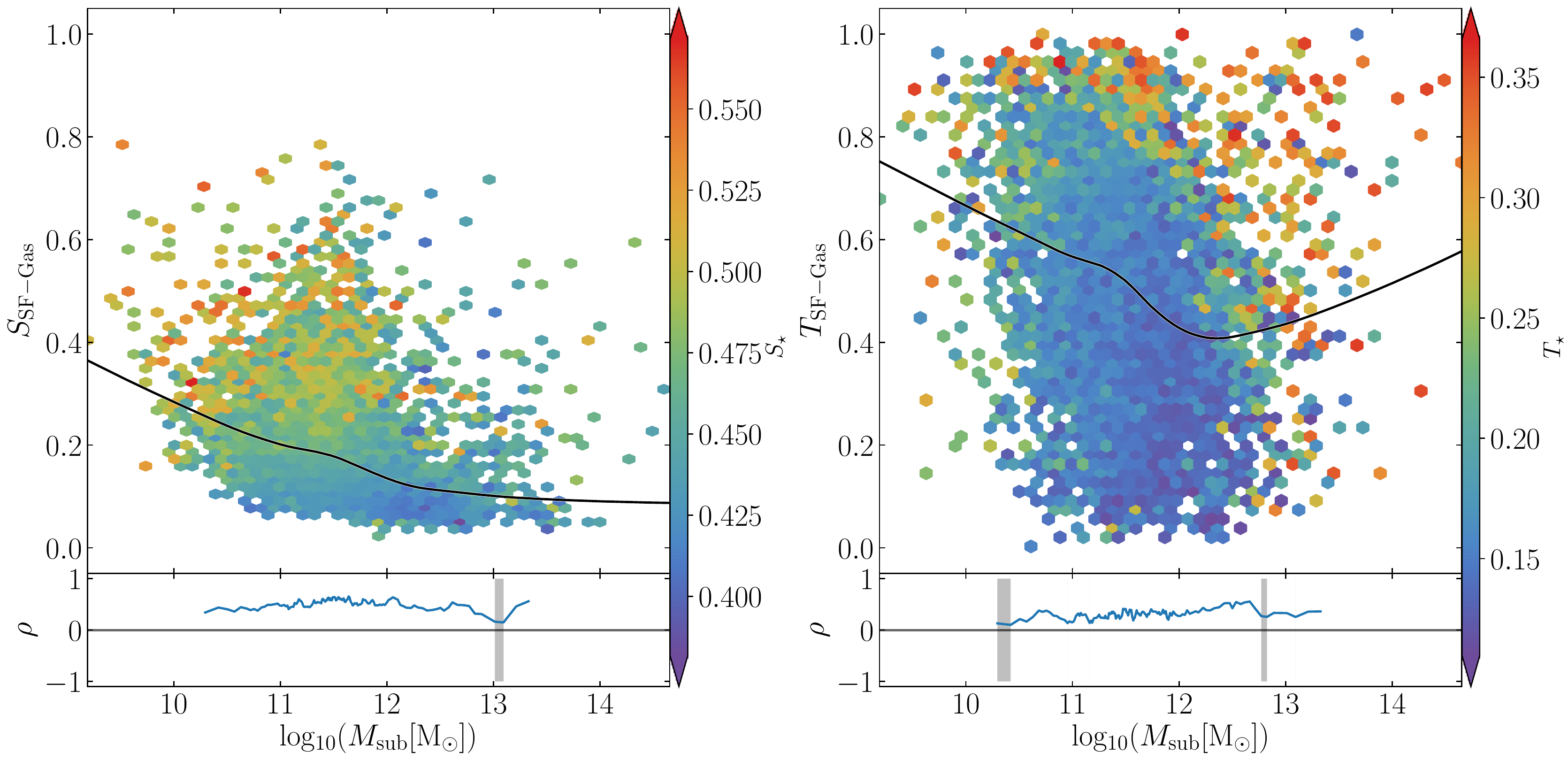}
\caption{
The sphericity (left-hand panel) and triaxiality (right-hand panel) of the star-forming gas as a function of subhalo mass. Black curves denote the running medians of the shape parameters, computed via the locally-weighted scatterplot smoothing method (LOWESS). Colours represent the median sphericity/triaxiality of the \textit{stellar} mass distributions of the subhaloes represented by each hexbin. The lower panels display the running Spearman rank correlation coefficient, $\rho$, between the residual shape parameters about $M_{\rm sub}$ for the two matter distributions, i.e. $\Delta S_{\rm SF-gas} - \Delta S_\star$ and $\Delta T_{\rm SF-gas} - \Delta T_\star$, where for instance $\Delta S_{\star, i} = S_{\star, i} - \tilde{S}_{\star}(M_{\mathrm{sub}, i})$. Grey shaded regions indicate mass ranges for which the correlation is recovered at low significance ($p > 0.05$). The subpanels corroborate, quantitatively, the impression given by visual inspection, that the morphology of the star-forming gas component correlates positively with that of the stellar component.}
\label{fig:shape_correlation}
\end{figure*}

We noted in the Section \ref{sec:shape_by_mass} that the star-forming gas configuration in massive subhaloes is often well described by a flattened prolate spheroid, similar to the characteristic morphology of the stars and DM in such structures. \citet{thob19} previously demonstrated that EAGLE galaxies with flattened stellar distributions are preferentially hosted by flattened DM haloes, motivating a closer examination here of the degree to which the morphology of the star-forming gas correlates with that of the other matter components. Since the density of stars typically dominates over the density of dark matter within the region traced by the star-forming gas, we focus on the correspondence between the morphology of the star-forming gas and the stars.

The main panels of Fig. \ref{fig:shape_correlation} show, as a function of subhalo mass, the sphericity (left-hand panel) and triaxiality (right-hand panel) of the star-forming gas distributions of the subhaloes comprising our sample. The distribution is shown as a 2D histogram, and black curves denote the running median of the star-forming gas shape parameters, computed via the locally weighted scatterplot smoothing method \citep[LOWESS; e.g.][]{cleveland}. The LOWESS curves are plotted within the interval for which there are at least 10 measurements at both lower and higher $M_{\rm sub}$. The colour of each hexbin denotes to the median value of the corresponding shape parameter of the \textit{stellar} component: subhaloes in bins denoted by red (blue) colours typically have a stellar component with a high (low) value of the shape parameter in question. 

The shape parameters of the star-forming gas and stellar distributions are strongly and positively correlated at effectively all subhalo masses: flattened star-forming gas distributions are generally found in subhaloes with flattened stellar components, and more prolate star-forming gas distributions are found in subhaloes with more prolate stellar components. We quantify the strength and significance of these correlations by computing a `running' Spearman rank correlation coefficient, $\rho(M_{\rm sub})$, for the $\Delta S_{\mathrm{SF-gas}} - \Delta S_{\star}$ and $\Delta T_{\mathrm{SF-gas}} - \Delta T_{\star}$ relations, where $\Delta X_{\mathrm{m}}$ represents the residual of shape parameter $X$ for matter distribution $m$ about the LOWESS median. Hence, in the case of sphericity, $\Delta S_{\star, i} = S_{\star, i} - \tilde{S}_{\star}(M_{\mathrm{sub}, i})$ for the $i^{\rm th}$ subhalo. The running Spearman rank correlation coefficient is computed in subhalo mass-ordered subsamples: for bins with a median subhalo mass $M_{\mathrm{sub}} < 10^{12.5}\mathrm{M_{\odot}}$, we use samples of 200 subhaloes with starting ranks separated by 50 subhaloes (e.g. subhaloes 1-200, 51-250 and 101-300). For bins with median $M_{\rm sub} > 10^{12.5}\mathrm{M_{\odot}}$, we use samples of 50 subhaloes with starting ranks separated by 25 subhaloes, to ameliorate the effect of the relative paucity of massive subhaloes. This running $\rho(M_{\rm sub})$ is plotted in the lower subpanel. Regions shaded in grey denote a Spearman rank $p$-value is $>0.05$, and thus indicate where the recovered correlation cannot be considered significant. 

The relatively high correlation coefficient ($\rho \gtrsim 0.4$) for the sphericity over a wide range in subhalo mass indicates that the degree of flattening of the two components is indeed strongly and positively correlated. The correlation is weaker ($\rho \gtrsim 0.3$) for the triaxiality parameter, but remains positive and significant over a wide range of subhalo masses. We have also examined the correlation of the shape parameters of star-forming gas with those of their host subhalo's DM, and we find that the correlation is not formally significant at any subhalo mass.

Our results suggest that examination of a large sample of galaxies with high-fidelity radio imaging is likely to reveal significant correlations between the radio continuum and optical morphologies of galaxies. There is not currently a firm consensus amongst observational studies, which are necessarily limited to comparisons of \textit{projected} ellipticities, in regard to correlations between the morphologies of the radio continuum and optical components of galaxies. \citet{battye_and_browne_09} report a strong, positive correlation of the two in late-type galaxies, and a weak negative correlation for early-type galaxies, whilst complementary studies using a smaller sample \citep{patel10}, or a sample of fainter, more-distant galaxies \citep{tunbridge}, recovered no significant correlations. More recently, \citet{hillier_19} examined the correlation of optical and radio continuum measurements of shape and orientation for galaxies in the COSMOS field, and recovered a significant correlation of position angles (projected orientation) between matched 3 GHz radio (VLA) and optical (\textit{HST}-ACS) images (seen in their Figs.~5 and 6).

\begin{figure*}
\centering
\hspace{-0.2cm}
     \includegraphics[width = 0.9\textwidth]{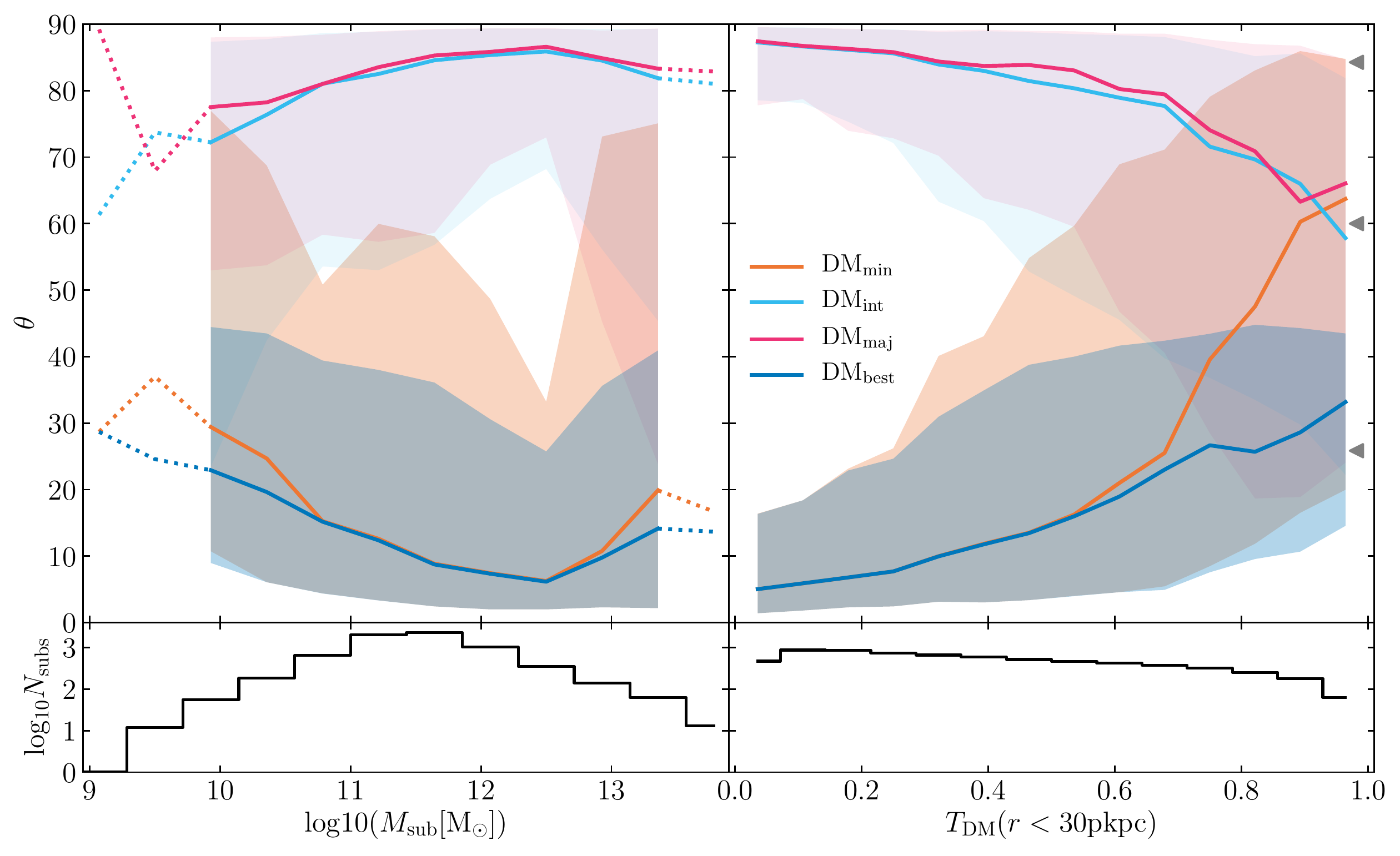}
\caption{The misalignment angle, $\theta$, of the minor axis of star-forming gas of our sample of subhaloes with respect to each of the principal axes of their DM distribution. The solid lines indicate the binned median values of $\theta$, whilst the shading denotes the 10th - 90th percentiles. The values are shown as a function of subhalo mass ($M_{\rm sub}$, left-hand panel) and DM triaxiality ($T_{\rm DM}$, right-hand panel). The orange, cyan and magenta curves correspond to the alignment of the star-forming gas minor axis with respect to the minor, intermediate and major axis of the DM, respectively, whilst the blue curve is with respect to the principle DM axis with which the star-forming gas minor axis is most closely aligned. The dotted lines indicate where the sampling drops below 30 subhaloes per bin. Subpanels show the number of subhaloes per bin. The grey arrows denote the expected 10th, 50th and 90th percentile values of alignment angles between vectors randomly oriented in 3 dimensions. In general the minor axis of star-forming gas is well aligned with the minor axis of its corresponding DM, but in strongly triaxial subhaloes the former often aligns more closely with the intermediate or major axis of the DM distribution.}
\label{fig:best_align}
\end{figure*}

\section{The alignment of star-forming gas with galaxies and their DM haloes}
\label{sec:results_alignments}

In this section we examine the orientations of the 3D distribution of star-forming gas in galaxies with respect to the stellar and DM components of their host subhaloes. We begin in Section \ref{sec:morphological_alignments} with an examination of the morphological alignment of subhalo components as a function of subhalo mass, triaxiality and cosmic epoch. In Section \ref{sec:kin_morph_align} we consider the alignment of the morphological minor axis of the star-forming gas with its kinematic axis.

\subsection{Morphological alignment of subhalo matter components}
\label{sec:morphological_alignments} 

We quantify the morphological alignment of the various components via the angle, $\theta$, between the minor axes of the ellipsoids describing each matter distribution, such that $\theta = 0^{\circ}$ indicates perfect alignment and $\theta = 90^{\circ}$ indicates orthogonality. As noted in Section \ref{sec:redit}, we consider the minor axis to be the natural choice when focusing on discs, as the minor axis is the most distinct axis for oblate discs (though we reiterate the finding from Section \ref{sec:shape_by_mass} that many flattened star-forming structures are mildly prolate). Moreover, as seen in Section \ref{sec:shape_by_mass}, the central regions of the stellar and DM distributions (to which the iterative reduced mass distribution tensor is more strongly weighted) also tend to be mildly oblate. 

Fig. \ref{fig:best_align} shows the alignment between the star-forming gas distribution and that of the DM. In the left-hand panel the alignment is shown as a function of subhalo mass ($M_{\rm sub}$) and in the right-hand panel it is shown as a function of the triaxiality of the DM. The thick orange curve and associated shading denotes the median alignment angle, and the 10th-90th percentiles of the distribution, when considering the minor axes of the two components. In general, the alignment is strong, with the median alignment angle typically $\simeq 30^\circ$ for $M_{\rm sub}=10^{10}\Msun$, declining to $\simeq 15^\circ$ for $M_{\rm sub}=10^{11}\Msun$ and $\simeq 10^\circ$ for $M_{\rm sub}=10^{12-12.5}\Msun$. In more massive subhaloes, the characteristic alignment is typically (marginally) poorer, rising to $\simeq 20^\circ$ for $M_{\rm sub}\gtrsim 10^{13}\Msun$. 

Examination of the right-hand panel shows that the alignment of the minor axes of the star-forming gas and the DM of its host subhalo is a strong function of the latter's triaxiality, with oblate subhaloes exhibiting close alignment of the two components ($\theta < 10^\circ$ for $T_{\rm DM} \lesssim 0.4$) but prolate subhaloes exhibiting much poorer alignment ($\theta > 50^\circ$ for $T_{\rm DM} \gtrsim 0.8$). As is clear from the scatter about the median relation, in prolate systems the minor axes of the two components can become effectively orthogonal. If the shape parameters of the two components are dissimilar, as is the case for the common configuration of an oblate disc within a prolate subhalo, alignment of the minor axes might not be the most likely scenario, since in such cases the minor and intermediate DM axes are not distinct. Indeed, the axes that should be `expected' to align are likely to be those most closely aligned with the angular momenta of the respective components (as we discuss in Section \ref{sec:kin_morph_align}). We therefore examine whether this is a genuine misalignment, or is rather a consequence of the minor axis of the star-forming gas exhibiting a preference to align with one of the other principal axes of the DM.

In Fig~\ref{fig:best_align}, the intermediate and major axes are denoted by the light blue and red curves, respectively. The dark blue curve denotes the angle between the minor axis of the star-forming gas and that of the DM axis with which it best aligns. In prolate subhaloes, for which the alignment quantified by the standard measure is poor, one can often find good alignment between the star-forming gas minor axis with one of the other principle axes of the DM. However, for $T_{\mathrm{DM}}(r<30\mathrm{pkpc})\rightarrow 1.0$ subhaloes, the characteristic alignments of the star-forming gas minor axis with all of the DM morphological axes converge towards $\simeq 60^\circ$, the expectation value for the alignment angle of unit vectors randomly oriented in 3 dimensions. This implies that poor alignment between the minor axes of the two components within high-triaxiality subhaloes is not primarily due to a preference for the star-forming gas minor axis to align with a non-minor DM morphological axis. Therefore in what follows, we focus exclusively on the misalignment between the minor axes of the two matter components.

\begin{figure}
\centering
\hspace{-0.2cm}
     \includegraphics[width = 0.45\textwidth]{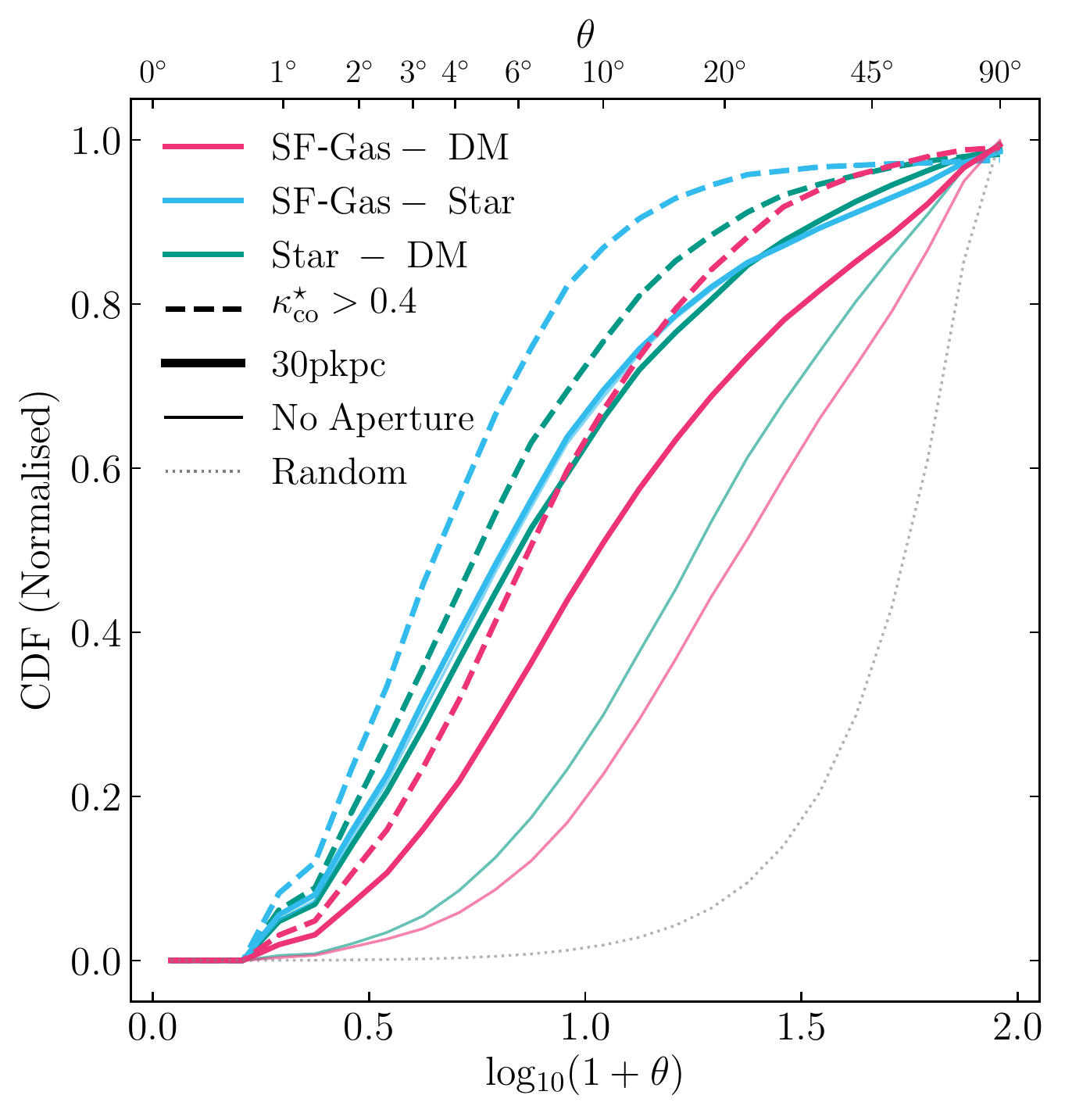}
\caption{The misalignment of the star-forming gas, DM and stellar distributions within the subhaloes of our sample. The figure shows the cumulative distribution function of the misalignment angle, $\theta$, between the minor axes of the matter distribution pairs in the legend. Thick curves correspond to fiducial measurements, thin curves denote alignments inferred when the initial characterisation of the mass distribution considers all particles of the relevant type bound to the subhalo, rather than only those within $30\pkpc$ of the subhalo centre. Thick dashed lines correspond to the subset of galaxies with $\kappa_{\rm co}^{\star}$, which broadly identifies star-forming disc galaxies. The dotted black line indicates the distribution of angles between randomly orientated vectors in 3D. Star-forming gas is a poorer tracer of the orientation of the subhalo DM distribution than are the stars.}
\label{fig:3D_align}
\end{figure}

Fig. \ref{fig:3D_align} shows the cumulative distribution function of the alignment angle $\theta$ for the three pairs of matter components, namely star-forming gas and DM (pink), star-forming gas and stars (blue), and stars and DM (green). We plot the distribution as a function of $\log_{10}(1+\theta)$ because the bulk of the misalignments (for all component pairs) are small, but there are long tails to severe misalignments. Thick lines denote our fiducial measurement, whilst the thin lines show the alignments inferred when the initial characterisation of the mass distribution considers all particles of the relevant matter component bound to the subhalo, rather than only those within $30\pkpc$ of the subhalo's centre. We show the latter in order to highlight the influence of the initial aperture, since an influence is to be expected: for example, \citet{vel15a} showed that the alignment of the stellar and DM components is stronger closer to the subhalo centre, i.e. that galaxies are best aligned with the local, rather than global, distribution of matter in the subhalo. For reference, the dotted black line shows the distribution function of alignment angles between randomly oriented vectors.

\begin{figure*}
\centering
\hspace{-0.2cm}
     \includegraphics[width = 0.45\textwidth]{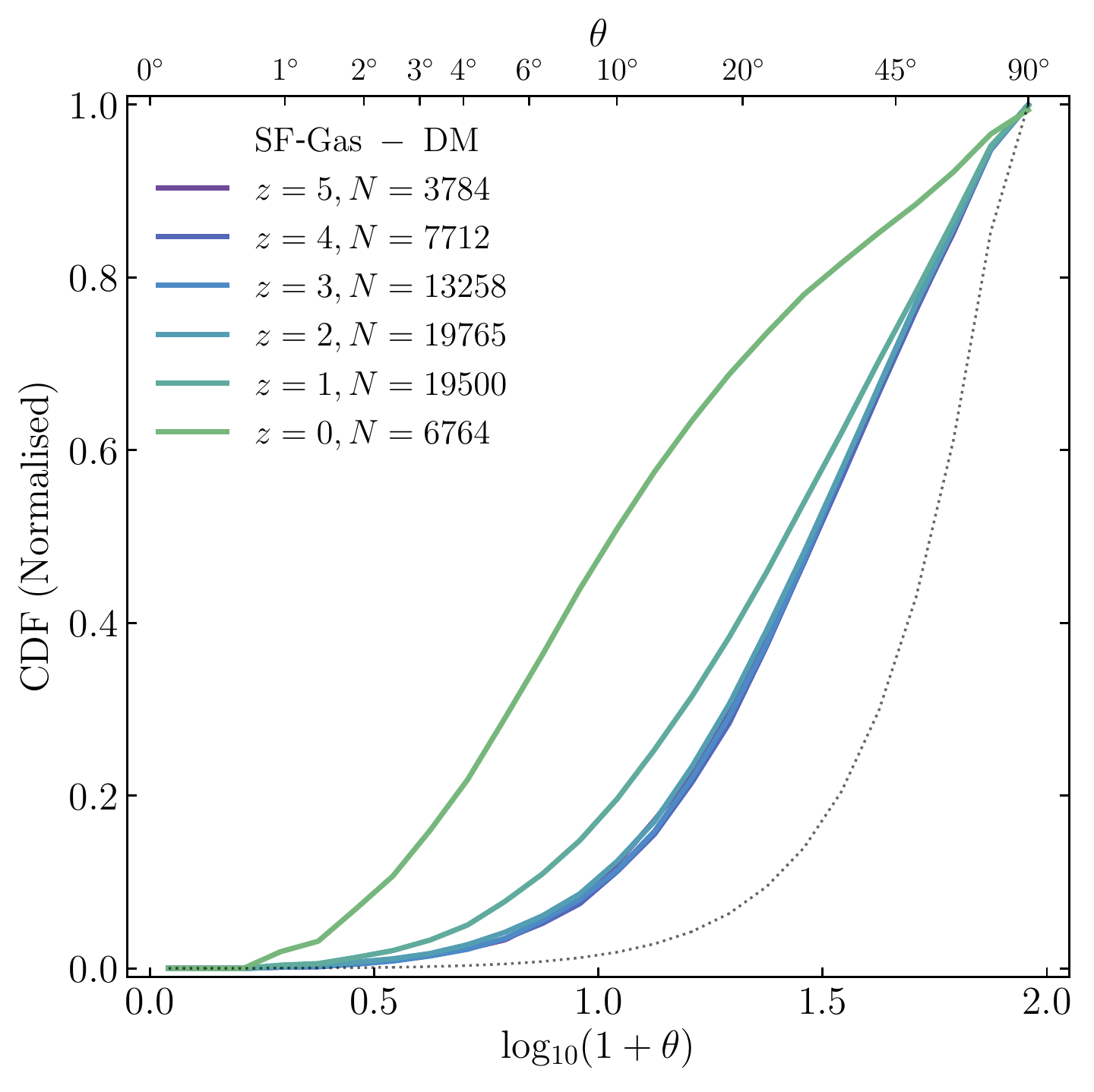}
     \includegraphics[width = 0.45\textwidth]{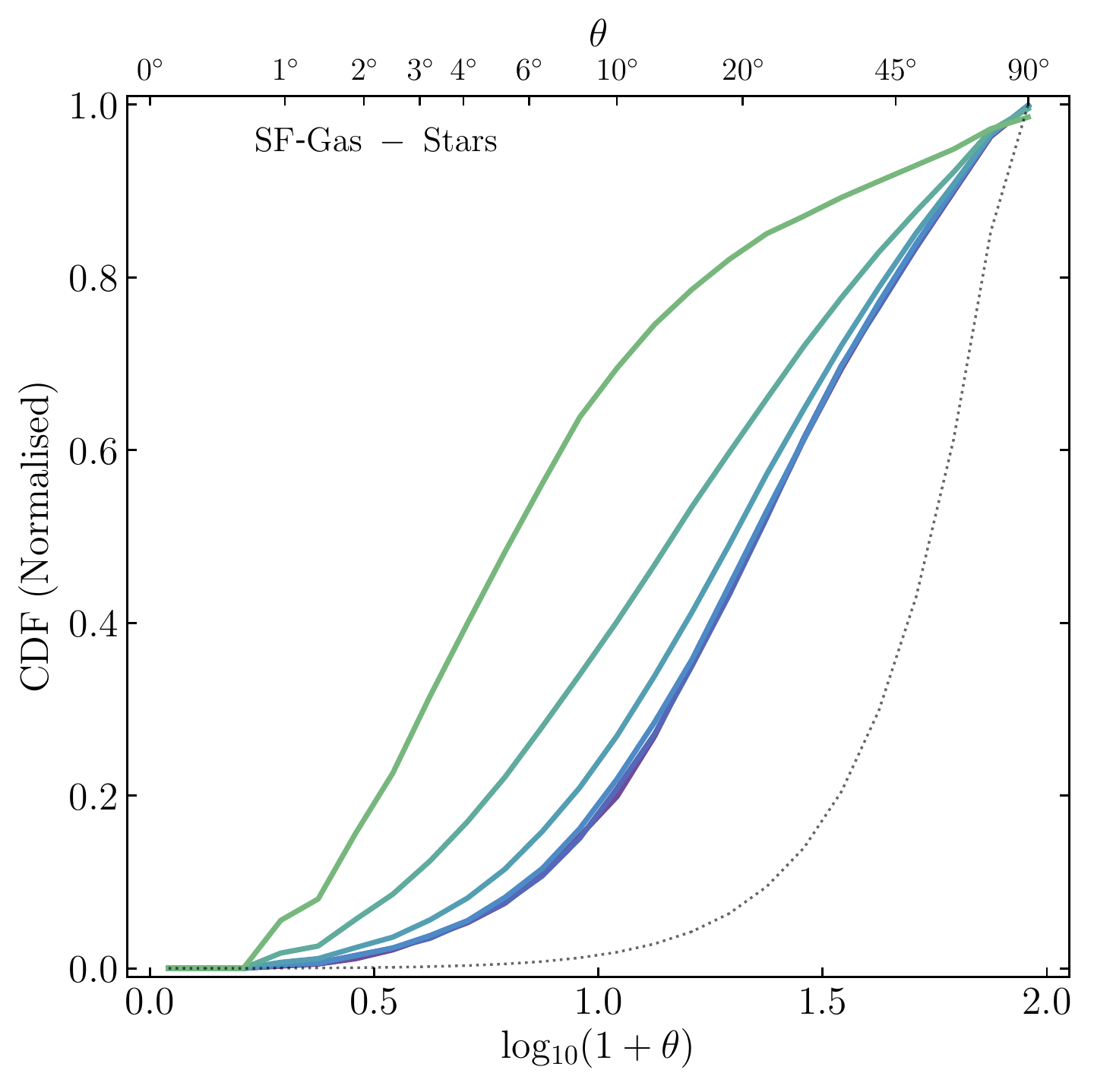}
\caption{The temporal evolution of the misalignment of the minor axes of the star-forming gas and DM distributions (left-hand panel) and of the star-forming gas and stellar distributions (right-hand panel) comprising the subhaloes of our sample, as computed while adopting a $30\pkpc$ aperture. The figure shows the cumulative distribution function of the misalignment angle, $\theta$. Colour indicates the redshift, and the thin black dotted line shows the distribution of angles between randomly orientated 3D vectors.}
\label{fig:3D_align_zdep}
\end{figure*}

For our fiducial measurements, half of the sampled subhaloes have star-forming gas distributions misaligned with their stellar components by more than $5^\circ$, and half have star-forming gas distributions misaligned with their DM component by more than $9.5^\circ$. Half of the subhaloes have stellar components misaligned with their DM component by more than $6^\circ$. Assessing the alignments recovered when considering all the particles of a given type associated with subhaloes, we find that half of the subhaloes have stellar-DM misalignments greater than $17^\circ$. The poorer star-forming gas - DM alignment with respect to the stars - DM alignment might be expected; since the stars and DM are collissionless components, their relevant evolutionary time-scale is the gravitational dynamical time, $t_{\rm dyn} = 1/\sqrt{G\rho} \sim 10^9\yr$, such that their morphologies and orientation effectively `encode' their formation and assembly history over an appreciable fraction of a Hubble time. In contrast, the phase-space structure of the collissional, dissipative gas is not preserved as it accretes onto galaxies and condenses into star-forming clouds. Its morphology and orientation therefore reflects a more instantaneous snapshot of the evolution of the subhalo than is the case for the collissionless components.

We note that the stellar - DM alignment shown in Fig.~\ref{fig:3D_align} (thick green curve) is significantly better than that inferred by \cite{vel15a}, who found that half of all the subhaloes they examined had misalignments worse than the $40^\circ$. This follows primarily from our use of an initial particle selection within a $30\pkpc$ sphere and the iterative reduced inertia tensor (which weights more strongly towards the halo centre), and also in part due to their measurement of the misalignment angle relative to the major axes of the mass distribution, and the slightly different sample selections. The influence of the initial particle selection can be assessed by comparison of the thick and thin solid curves: as expected, when one considers all matter bound to the subhalo (as opposed to only that within a $30\pkpc$ sphere) when initialising the iterative characterisation of the mass distribution, the misalignments with respect to the DM become significantly more pronounced. As is clear from the thinner curves of Fig. \ref{fig:3D_align}, in this case half of the sampled subhaloes have star-forming gas distributions misaligned with their DM components by more than $\simeq20^\circ$, and half have stellar components misaligned with their DM component by more than $15^\circ$. The misalignment of star-forming gas and the stars is however largely unaffected, since the bulk of both components is typically found within the central $30\pkpc$.

Having noted that misalignments are typically most severe in massive, prolate subhaloes, which tend to host quenched elliptical galaxies \citep[see e.g.][]{thob19}, it is reasonable to hypothesise that subhaloes hosting star-forming disc galaxies (i.e. those with $\kappa_{\rm co}^{\star}>0.4$) will exhibit significantly better alignment than the broader sample. The misalignment angles for this subset of subhaloes are shown by the dashed lines in Fig. \ref{fig:3D_align}, and indeed we find that the primary consequence of restricting our focus to these systems is the exclusion of galaxies with severe misalignments. For this subsample, only 20 percent of galaxies exhibit star-forming gas distributions misaligned with their DM by more than $\simeq 10^\circ$. 

Fig. \ref{fig:3D_align_zdep} shows the temporal evolution of the misalignment angle, $\theta$, of the minor axes of star-forming gas and DM mass distributions (left-hand panel) and the star-forming gas and stars (right-hand panel). Here, as was the case for Fig. \ref{fig:prog}, we consider at all epochs subhaloes that satisfy the selection criteria specified in Section \ref{sec:sample}, however we do not here focus solely on main branch progenitors of $L_{\star}$ subhaloes. It is immediately apparent that the orientation of the star-forming gas is a much poorer tracer of the orientation of both the DM and the stars at early cosmic epochs than at the present-day (though the characteristic alignment is always much better than random). As noted above, at $z=0$ half of the sampled subhaloes have star-forming gas distributions misaligned with their DM components by more than $10^\circ$, but at $z=1$ half are misaligned by more than $25^\circ$ and at $z=2$ the figure is $29^\circ$. Similarly, at $z=0$ half of the sampled subhaloes have star-forming gas distributions misaligned with their stellar components by more than $6^\circ$, but at $z=1$ half are misaligned by more than $14^\circ$ and at $z=2$ half are misaligned by at least $19^\circ$. The deterioration of the alignment of the star-forming gas distribution with both the DM and the stars at earlier times is to be expected, since all three components tend to be more spherical (less flattened) at higher redshift. Although in principle even highly spherical distributions can exhibit perfect alignment, as $S \rightarrow 1$ the minor axis becomes less well defined.

In Appendix \ref{sec:fits} we provide analytical fits to probability distribution functions of the misalignment angle, $\theta$, of the star-forming gas distribution with those of DM and stars, enabling subhaloes in dark matter-only simulations to be populated with galaxies whose star-forming gas has a realistic misalignment distribution. 

\subsection{Alignment of the kinematic and morphological axes}
\label{sec:kin_morph_align}

\begin{figure}
\centering
\hspace{-0.2cm}
     \includegraphics[width = 0.45\textwidth]{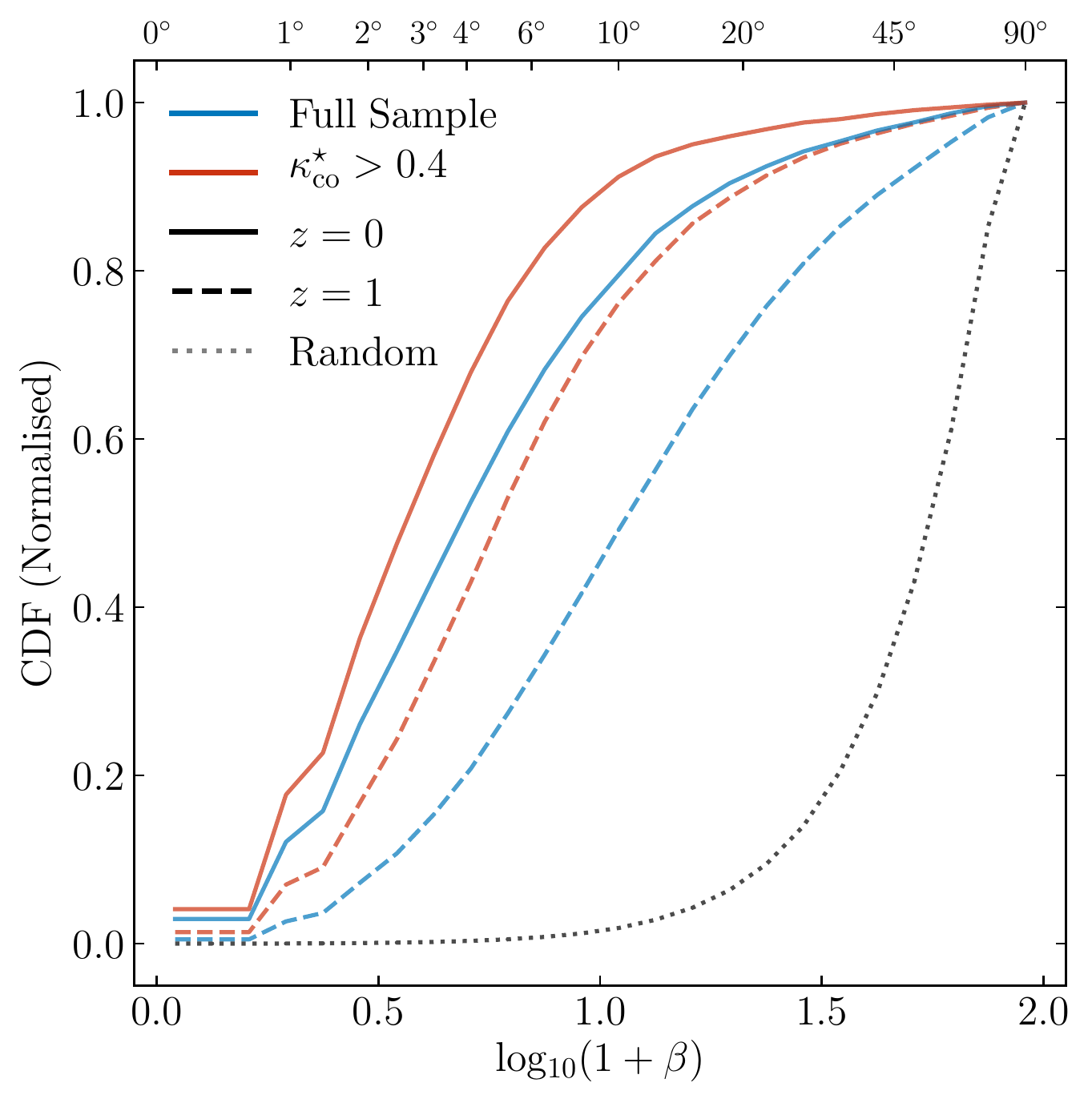}
     \caption{A histogram of the alignment between the morphological and kinematic axes of the star-forming gas within our sample. Alignment angles are given in terms of $\mathrm{log_{10}(1 + \beta)}$, as the majority of the subhaloes have small $\beta$. The solid (dashed) lines correspond to subhaloes at redshift $z = 0(1)$. The black dotted line shows the distribution of angles between randomly orientated 3D vectors. Red lines correspond to the subset of galaxies with $\kappa_{\rm co}^{\star}$, which broadly identifies star-forming disc galaxies, while blue lines correspond to the full sample.}
\label{fig:sfg_minor_kin}
\end{figure}

A novel aspect of radio continuum lensing surveys is that complementary observations of the 21cm hyperfine transition emission line from atomic hydrogen can, in principle, be obtained simultaneously with little or no extra observing time. The Doppler shift of the 21cm line is widely used to infer the kinematics of the atomic phase of the ISM \citep[e.g.][]{bosma_78,swaters_99} and hence affords an independent means of assessing galaxy orientation. As noted by \citet{blain}, \citet{morales} and \citet{deburghday}, the kinematic axis can be used as a proxy for the unsheared morphological axis, and hence affords a means to suppress the influence of galaxy shape noise and intrinsic alignments. 

Clearly, the na\"ive application of this method assumes perfect alignment of the kinematic and minor morphological axes. To assess the accuracy of this assumption, we define the morphokinematic misalignment angle, $\beta$, as the angle between the minor axis of the star-forming gas distribution, and the unit vector of its angular momentum. Fig. \ref{fig:sfg_minor_kin} shows the cumulative distribution function of $\log_{10}(1+\beta)$, with solid curves denoting present-day measurements and dashed lines denoting measurements at $z=1$. The blue curves correspond to the fiducial sample, whilst red curves correspond to the subset of galaxies with $\kappa_{\rm co}^{\star} > 0.4$. For reference, the dotted black line again shows the distribution function of alignment angles between randomly oriented vectors.

As na\"ively expected, the star-forming gas minor axis and angular momentum vector of star-forming gas are well aligned for present-day subhaloes: 80 percent of systems exhibit morphokinematic misalignments of less than $10^\circ$. However, similar to the internal component alignments, the distribution function exhibits a long tail to severe, but rare, misalignments. The morphokinematic alignment improves if one restricts the analysis to the $\kappa_{\rm co}^\star >0.4$ subsample, for which eighty percent of the systems are misaligned by less than $6^\circ$, and the tail to severe misalignments is strongly diminished. At might be expected when considering the reduced prevalence of strongly flattened star-forming discs at $z=1$, the morphokinmatic alignment is poorer at this earlier epoch, with 80 percent of subhaloes aligned to better than $30^\circ$, and $12^\circ$ when restricting to the $\kappa_{\rm co}^\star >0.4$ subsample.

To establish the characteristics of the subhaloes that typically suffer from poor morphokinematic alignment, we separate the primary sample of present-day subhaloes into quartiles of $\beta$, and quote in Table \ref{tab:kinematic} the median values of key characteristics of subhaloes in each quartile, namely the star-forming gas sphericity, subhalo mass, star formation rate, stellar mass, the star-forming gas corotation parameter and the half-mass radius of the star-forming gas. This exercise illustrates that poor alignment of the minor axis of the star-forming gas with its angular momentum vector is more typical in subhaloes hosting a spheroidal central galaxy, with a low star-formation rate and a less flattened and less extended star-forming gas distribution. In principle, such systems can be readily identified from either optical or radio continuum imaging.

\begin{table}

\begin{tabular}{l|cccc}

Quartile & $\mathrm{1^{st}}$ & $\mathrm{2^{nd}}$ & $\mathrm{3^{rd}}$ & $\mathrm{4^{th}}$
\\
\hline \\
$S_{\mathrm{SF-gas}}$ & 0.11$\pm$0.05 & 0.14$\pm$0.06 & 0.19$\pm$0.07 & 0.29$\pm$0.12\\
$M_{\mathrm{sub}}$ & 11.67$\pm$0.5 & 11.53$\pm$0.52 & 11.41$\pm$0.51 & 11.44$\pm$0.67\\
SFR & 0.59$\pm$1.05 & 0.41$\pm$0.93 & 0.32$\pm$1.31 & 0.31$\pm$0.94\\
$M_{\star}$ & 9.97$\pm$0.54 & 9.75$\pm$0.56 & 9.6$\pm$0.52 & 9.56$\pm$0.61\\
$\kappa_{\rm co}^{\rm SF}$ & 0.93$\pm$0.06 & 0.9$\pm$0.08 & 0.86$\pm$0.1 & 0.76$\pm$0.18\\
$r_{\mathrm{SF-gas}}$ & 8.29$\pm$5.37 & 5.88$\pm$6.74 & 3.96$\pm$9.47 & 2.58$\pm$29.75

\end{tabular}
\caption{The median and standard deviation in various subhalo properties for the systems binned into quartiles based on the alignment angle between the star-forming gas kinematic and morphological minor axes. In degrees, the 0$^{\mathrm{th}}$, 25$^{\mathrm{th}}$, 50$^{\mathrm{th}}$, 75$^{\mathrm{th}}$ and 100$^{\mathrm{th}}$ percentile values are 0.0$^{\circ}$, 2.14$^{\circ}$, 4.36$^{\circ}$, 9.15$^{\circ}$ and 89.75$^{\circ}$ respectively. The values from top to bottom are: the star-forming gas sphericity; total subhalo mass (as $\mathrm{log_{10}} M_{\mathrm{sub}} [\mathrm{M}_{\odot}]$); star-formation rate within  $30\pkpc$ (in $\mathrm{M_{\odot}yr^{-1}}$); stellar mass within  $30\pkpc$ (as $\mathrm{log_{10}} M_{*} [\mathrm{M}_{\odot}]$); the fraction of kinetic energy in the star-forming gas invested in corotation; and the star-forming gas half-mass radius (in pkpc).}
\label{tab:kinematic}

\end{table}

\section{The morphology and alignment of projected star-forming gas distributions}
\label{sec:results_2d}

In this section we examine the morphologies, alignments and orientations of star-forming gas and DM when projected `on the sky' in 2 dimensions, affording a direct connection with observational tests. In Section \ref{sec:proj} we consider the ellipticity of the matter components, i.e. their projected morphology. In Section \ref{sec:proj_align} we consider the projected alignments of galaxies. 

\subsection{Projected ellipticities}
\label{sec:proj}

\begin{figure}
\centering
\hspace{-0.2cm}
     \includegraphics[width = 0.45\textwidth]{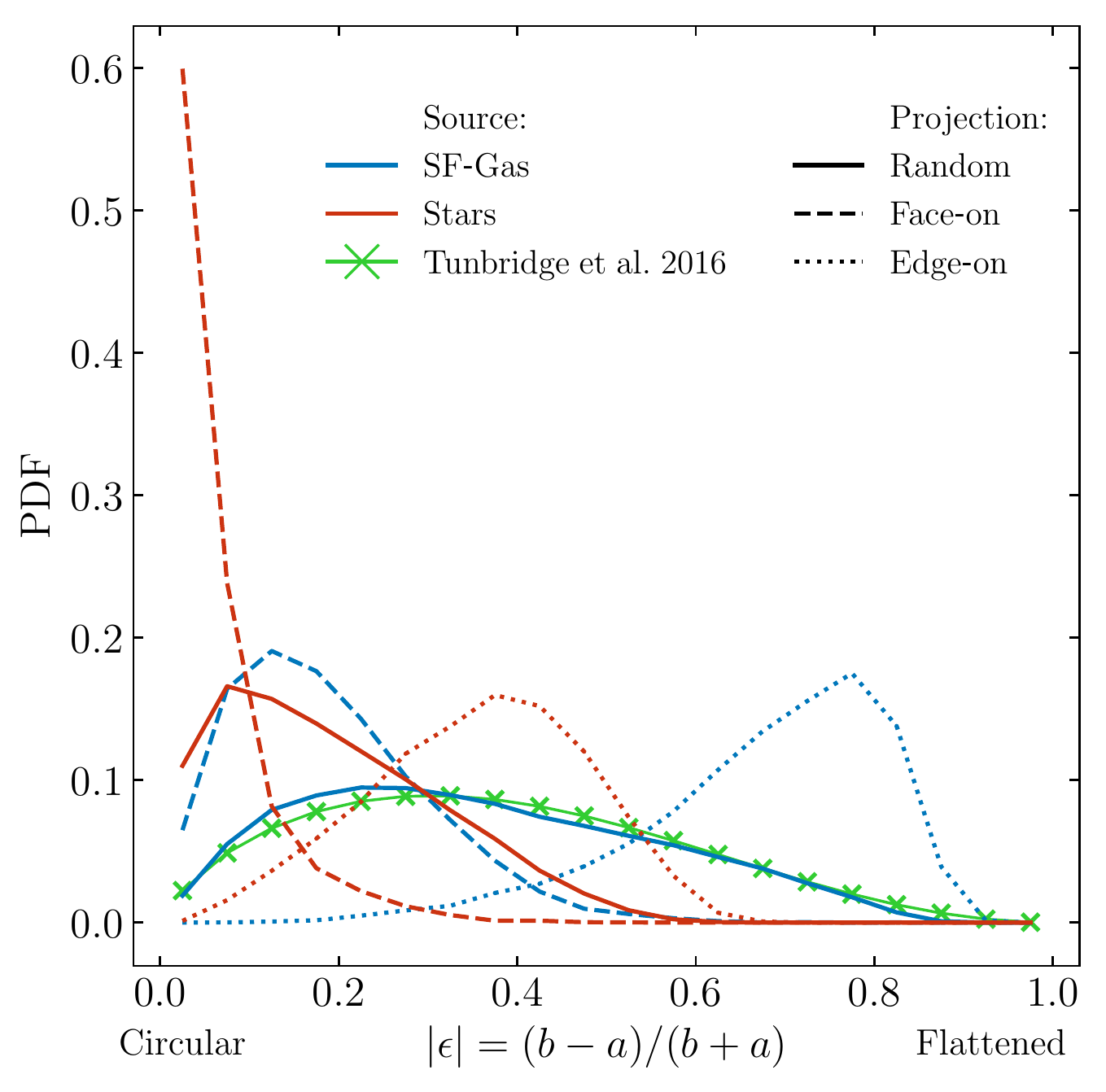}
\caption{Probability distribution functions of the projected 2D ellipticities of the present-day mass distributions of star-forming gas (blue curves) and stars (red curves) bound to the subhaloes of our sample. The solid curves denote the aggregated ellipticities recovered from projection of the 3D mass distributions along 100 random axes of projection. For reference, the dashed and dotted curves show the distributions recovered when the galaxies are oriented face-on and edge-on, respectively, to the axis of projection. An observational comparison (green curve with crosses) is sourced from \citet{tunbridge}, who provide a best-fitting model to the distribution of observed ellipticities of galaxies in the radio VLA COSMOS data set.}\label{fig:proj_view}
\end{figure}

It is via measurement of the morphology of galaxies in projection, i.e. their ellipticity, that the weak gravitational shear is estimated. Since galaxies are intrinsically ellipsoidal (i.e. non-circular), the observed ellipticity is due to both the intrinsic ellipticity of the galaxy, and the lensing shear. The former can therefore be considered as a noise term when measuring the shear, and is often referred to as `shape noise'. Since the variance of the observed ellipticity, $\epsilon_{\rm obs}$ is the sum of the variances of the intrinsic ellipticity and the (reduced) shear, i.e. $\sigma^2(\epsilon_{\rm obs}) = \sigma^2(\epsilon_{\rm int})+\sigma^2(\epsilon_{\rm sh})$, the signal-to-noise ratio of shear measurements is markedly sensitive to the diversity of the intrinsic ellipticity of the galaxy population being surveyed. 

To measure the intrinsic ellipticity of matter distributions, we adapt the iterative reduced inertia tensor algorithm presented in Section \ref{sec:redit} to consider only two spatial coordinates and so recover the best-fitting ellipse. The intrinsic ellipticity is then $\lvert\epsilon_{\mathrm{ int}} \rvert = (a-b)/(a+b)$, where $a,b$ are the major and minor axis lengths of this ellipse, respectively, such that low ellipticity corresponds to near-circular morphology, and high ellipticity corresponds to a strongly flattened configuration. Hereafter we omit the subscript for brevity, such that $\epsilon \equiv \epsilon_{\rm{int}}$. As noted in Section~\ref{sec:redit}, the first iteration of the algorithm considers all particles of the relevant type within a circular aperture of radius $r_{\mathrm{ap}} = \mathrm{max}(30\pkpc, 2r_{\mathrm{sfg}})$, where $r_\mathrm{sfg}$ is the 2D half-mass radius of star-forming gas within a circular aperture of $30\pkpc$. The use of this additional criterion ensures a robust morphological characterisation of the image projected by the most extended gas discs when viewed close to a face-on orientation. At each iteration, the elliptical aperture adapts to maintain a constant area. 

Fig.~\ref{fig:proj_view} shows the probability distribution function of the projected ellipticity of star-forming gas (blue curves) and the stars (red curves) associated with the subhaloes of our sample. The solid curves denote the distribution of aggregated ellipticities recovered from projection of the 3D mass distributions along the line-of-sight of 100 `observers' randomly positioned on a unit sphere, thus crudely mimicking a real light cone (albeit without noise or degradation from instrumental limitations). The dashed and dotted curves show the ellipticity distributions recovered when the subhaloes are first oriented such that the projection axis is parallel to, respectively, the minor and major principal axes of the respective 3D mass distribution, in order to show the ellipticities when viewed face-on and edge-on. 

The distribution of ellipticities when projected along random lines of sight is significantly broader for the star-forming gas than is the case for the stars: the IQRs of two distributions are $0.30$ and $0.19$, respectively. The origin of this difference is revealed by inspection of the ellipticity distributions for the face-on and edge-on reference cases: as might be inferred from the distribution of 3D shape parameters, star-forming gas is more commonly found in flattened configurations (corresponding to large values of the projected ellipticity) than is the case for the stars. The characteristic ellipticity of the flattened structures is greater for the star-forming gas, as can be quantified via the median ellipticities, $\tilde{\epsilon}^{\rm sfg}_{\rm edge} = 0.70$ and $\tilde{\epsilon}^{\rm stars}_{\rm edge} = 0.37$. Consequently, when projected along random lines of sight, there is a mild but significant paucity of low-ellipticity star-forming structures, since observing such a configuration requires that the galaxy is oriented close to face-on. Similarly, there are few high-ellipticity stellar structures, but this deficit is greater: not only does observing such a configuration require that the galaxy is oriented close to edge-on but, crucially, stellar structures that are strongly flattened (in 3D) are rare. We note that our sample selection criteria act to \textit{minimise} these differences, since galaxies with significant star-forming gas reservoirs preferentially exhibit flattened stellar discs, i.e. elliptical and spheroidal galaxies are under represented by our sample. 

The solid green curve of Fig.~\ref{fig:proj_view} denotes the best-fitting functional form of the galaxy ellipticity distribution recovered from the application of the \textsc{im3shape} algorithm \citep{zuntz} to Very Large Array (VLA) $L$-band observations of galaxies in the COSMOS field \citep[][see their equation 8]{tunbridge}. The iterative algorithm finds the best-fitting two-component Sèrsic (disc and bulge) model, yielding two-component ellipticities $\mathbf{\epsilon}=(e_1,e_2)$, and is similar in concept, if not in detail, to the approach used here to characterise the simulated galaxies. There is a  remarkable correspondence between the observed ellipticity distribution and that recovered from EAGLE. The qualitative similarity is a reassuring indication that the ellipticity distribution of star-forming gas yielded by EAGLE is realistic, however we caution that the degree of agreement is likely to be, in part, coincidental: besides the differences in shape measurement algorithms and the absence of noise or smearing by a point spread function in the simulated shape measurements, the observed sample also spans a wide range of redshifts. 

\begin{figure}
\centering
\hspace{-0.2cm}
     \includegraphics[width = 0.47\textwidth]{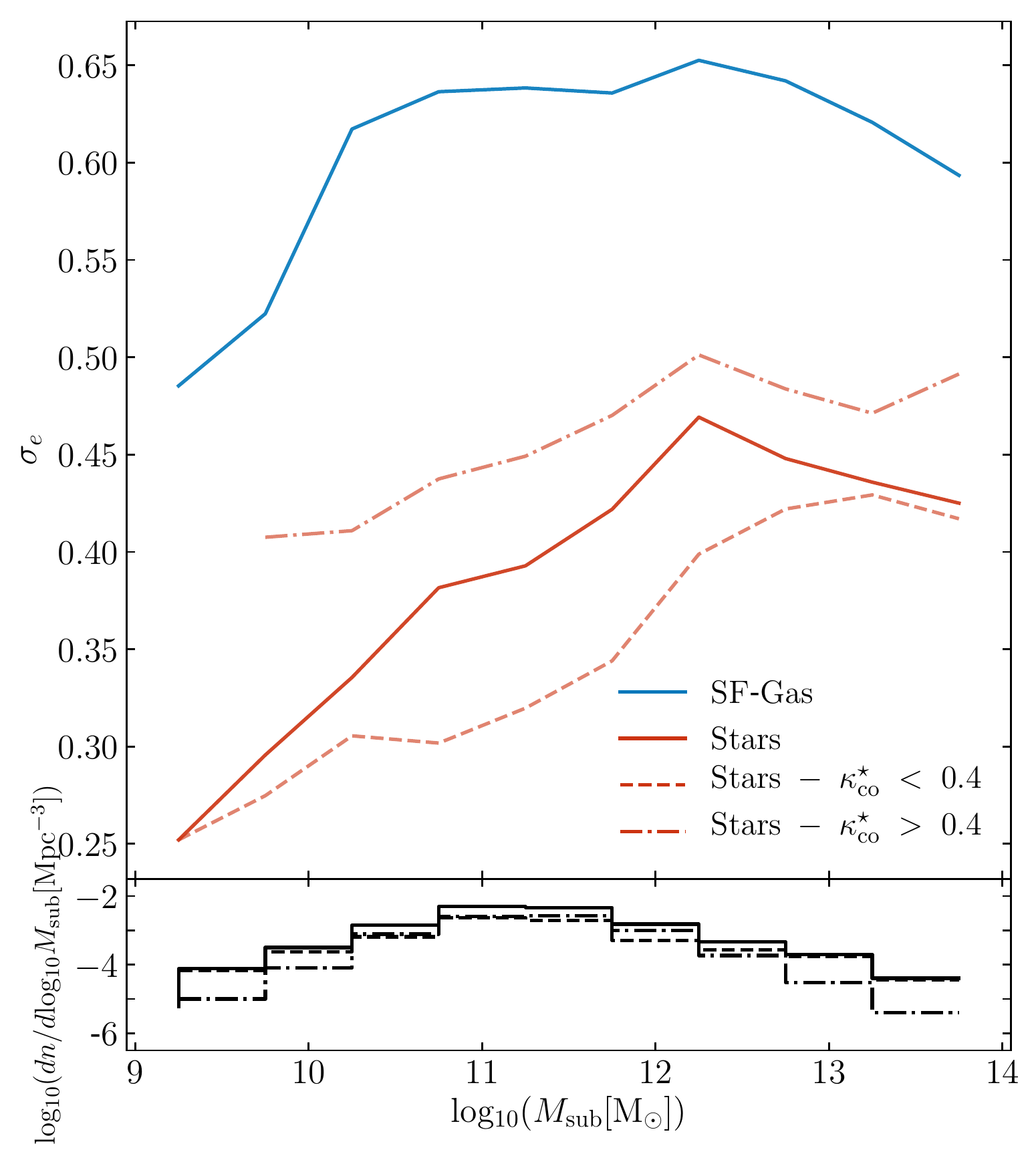}
\caption{The scatter in the projected ellipticities of stars and star-forming gas, calculated as the standard deviation of the minor-major axis ratios within a given mass bin. Blue lines relate to the star-forming gas, and red lines to the stars. The solid lines correspond to the shape error in the projected galaxy shapes and the dotted line corresponds to the edge-on star-forming gas particle coordinate projection for the star-forming gas. The red dashed and dot-dashed lines correspond to the standard deviations for $\kappa_{\mathrm{co}}^{\star} < 0.4$ and $\kappa_{\mathrm{co}}^{\star} > 0.4$ stellar systems respectively. $\kappa_{\mathrm{co}}^{\star}$ is here the fraction of kinetic energy invested in corotation for the \textit{stars}, as opposed to the star-forming gas, within  $30\mathrm{pkpc}$, as outlined by \citet{thob19}. The lower panel displays the volume density of subhaloes in a given mass bin, given as $\mathrm{log_{10}(dn/dlog_{10}M_{sub}[Mpc^{-3}])}$. At all masses, the shape noise is systematically greater for galaxy populations that are intrinsically flatter.}\label{fig:shape_noise}

\end{figure}

\citet[][]{tunbridge} noted that the dissimilar diversity of the projected ellipticities of the star-forming gas and stellar mass distributions is of practical relevance, because it governs the shape noise. This difference is analogous to the difference in shape noise in the optical regime expected for samples of early- and late-type galaxies: \citet{joachimi_13} estimate that the former exhibit up to a factor of two less shape noise than the latter at fixed number. We assess the magnitude of this effect in EAGLE, by defining the shape noise of a sample of galaxies, $\sigma_e$, as
\begin{equation}
\sigma_e^2 = \frac{1}{N}\sum_i \lvert e_i \rvert^2,
\end{equation}
where $N$ is the total number subhaloes in the sample. The quantity in the summation is often referred to as the polarisation \citep[see e.g.][]{blandford91} and is defined as $\lvert e \rvert=(a^2-b^2)/(a^2+b^2)$. It is thus related to the ellipticity via $e=2\epsilon/(1+\lvert\epsilon\rvert^2)$. 

We compute $\sigma_e$ for the star-forming gas and stellar distributions of subhaloes as a function of subhalo mass. These measurements are shown in Fig. \ref{fig:shape_noise}. The solid curves denote measurements for the star-forming gas (blue) and stars (red) considering all subhaloes comprising our sample. To place the difference in shape noise between the two matter types into context, we also show the shape noise of the stellar component when splitting the main sample into two subsamples separated about $\kappa_{\rm co}^\star = 0.4$, thus broadly separating the main sample into late- and early-type galaxies. The shape noise of the star-forming gas associated with subhaloes of all masses probed by our sample is systematically greater than is the case for their stars, by $\Delta \sigma_e \simeq 0.19-0.25$, an offset comparable to the difference between the shape noise (at fixed subhalo mass) of the stellar component of subhaloes comprising our crudely defined early- and late-type subsamples. \citet{tunbridge} report a qualitatively similar offset of the shape noise of radio continuum sources relative to their optical images (see their table 3). 

\subsection{Projected alignment}\label{sec:proj_align}

\begin{figure}
\centering
\hspace{-0.2cm}
     \includegraphics[width = 0.45\textwidth]{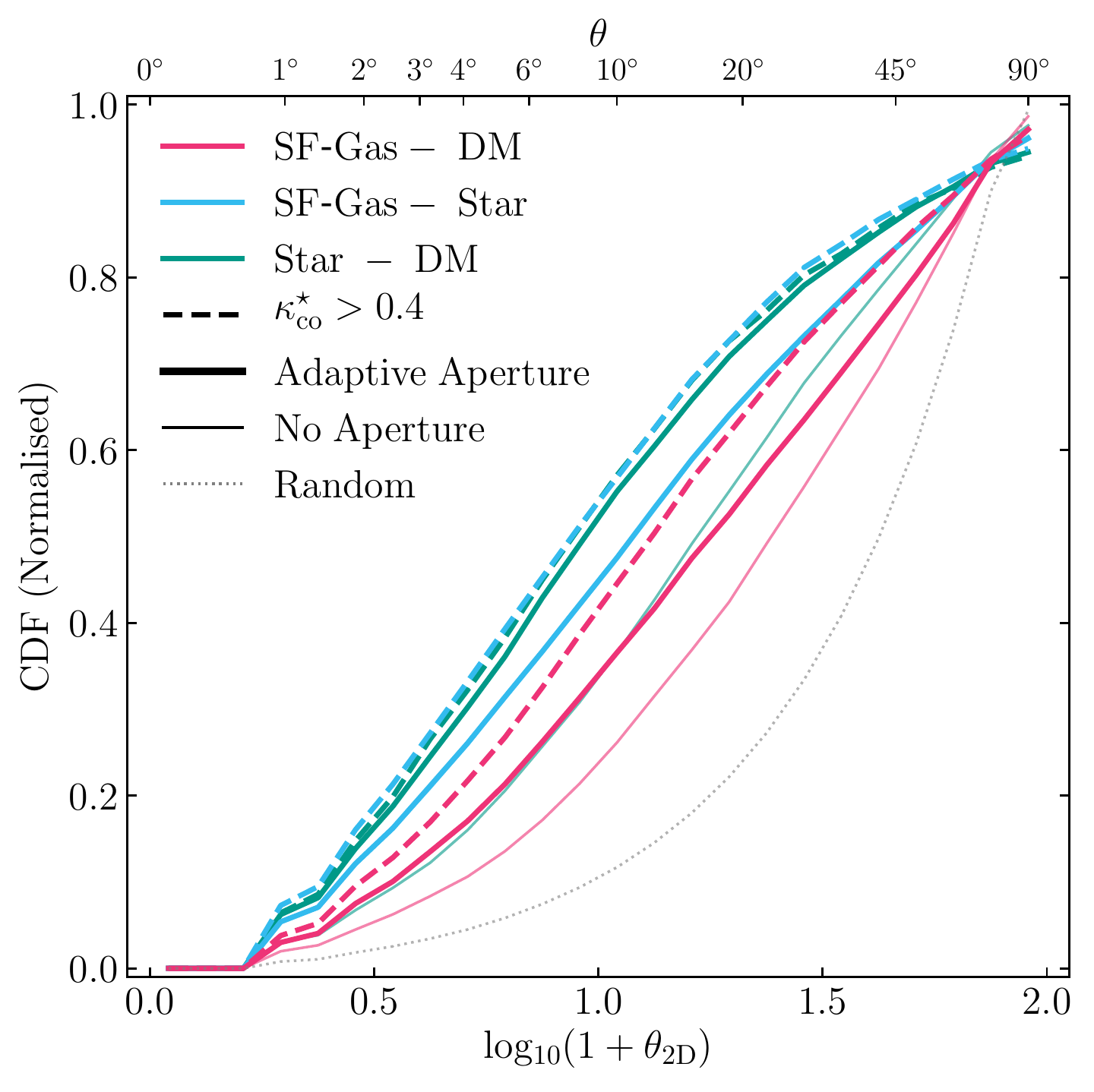}
\caption{The projected 2D internal alignment between the stars, DM and star-forming gas within the subhaloes of our sample. The figure displays a normalised cumulative distribution function of the angle $\theta_{\mathrm{2D}}$ between the minor axes of various matter distributions within subhaloes. The line colour indicates the two matter types assessed, thick dashed and thin solid lines correspond to the aperture used in the computation of the iterative reduced inertia tensor. The black dotted line indicates the distribution of angles between randomly orientated vectors in 2D. Star-forming gas is a poorer tracer of the underlying DM distribution than the stars in terms of orientation.}\label{fig:2D_align}
\end{figure}

In practice, it is only the misalignment angle of the various matter types \textit{in projection} that can be measured observationally. We therefore extend the exploration of 3D misalignments presented in Section \ref{sec:results_alignments}, to examine misalignments in projection. Fig. \ref{fig:2D_align} shows the cumulative distribution function of $\theta_{\mathrm{2D}}$, the alignment angle of the three pairs of matter components when viewed in projection. As with Fig. ~\ref{fig:3D_align}, we plot the distribution as a function of $\log_{10}(1+\theta_{\rm 2D})$ since the bulk of the misalignments are small, but show long tails to severe misalignments. Thick lines denote our fiducial measurement, whilst thin lines show the alignments inferred when the initial characterisation of the projected mass distribution considers all particles of the relevant matter component bound to the subhalo. Thick dashed lines repeat the fiducial measurement for the subsample of subhaloes hosting late-type galaxies, i.e. those with $\kappa_{\rm co}^\star > 0.4$. For reference, the dotted black line shows the distribution function of alignment angles between randomly oriented vectors.

The plot reveals that the projected alignments are qualitatively similar to those recovered in 3D, insofar that the star-forming gas and DM are most weakly aligned (half of all subhaloes are aligned to better than $16.9^\circ$), whilst the star-forming gas - stars and stars - DM alignments are aligned significantly more closely (half of all subhaloes aligned to better than $10.9^\circ$ and $8.1^\circ$, respectively). Discarding the initial aperture weakens the alignment between the more centrally concentrated baryons and the DM but, in a similar fashion to the 3D case, has little impact on the alignment between star-forming gas and stars. Restricting the sample to late-type galaxies improves the alignment of all component pairs, with half of all subhaloes being aligned to better than $12.2^\circ$, $7.7^\circ$  and $7.8^\circ$ for, respectively, the star-forming gas - DM, star-forming gas - stars, and stars - DM pairs. 

We note that, in contrast to \citet[][their Fig 10.]{tenneti14} and \citet[][their Fig. 13]{vel15a}, we find that the projected alignments are in general \textit{weaker} in projection than in 3D. For example, the median alignment angle of star-forming gas and DM using our fiducial aperture choices are $9.5^{\circ}$ in 3D and $16.9^{\circ}$ in projection. This is a consequence of our choice, motivated in Section \ref{sec:redit}, to measure misalignments relative to the minor axis rather than the major axis; whilst the projected misalignment is insensitive to this choice, the choice has a significant bearing on the alignments in 3D. We have explicitly confirmed that switching from the use of the minor axis to the major axis to define the misalignment angle results in smaller misalignments when projecting from 3D, consistent with the findings of \citet{tenneti14} and \citet{vel15a}. Although not shown in the figure, we have further examined the misalignment angles of all matter component pairs at $z=1$, and find more severe misalignments at the earlier cosmic epoch. This result is largely insensitive to the use of the axisymmetry criterion.

\section{Summary and Discussion}\label{sec:Discussion}

We have investigated the morphology of, and mutual alignments between, the star-forming gas, stars and dark matter bound to subhaloes that form in the EAGLE suite of simulations \citep[][]{schaye15,crain15,mcalpine16}. Our study is motivated by the complementarity of weak lensing experiments conducted using radio continuum surveys with traditional optical surveys. While recent radio weak lensing studies were limited by low source densities \citep[see e.g.][]{tunbridge, hillier_19, Harrison_2020}, the next-generation Square Kilometer Array (SKA) radio telescope will be competitive with optical surveys at a higher characteristic redshift. In simulations like EAGLE, gas that has a non-zero star formation rate is a good proxy for gas that is bright in the radio continuum. EAGLE represents a judicious test-bed for an assessment of this kind, as the simulations were calibrated to ensure a good reproduction of the galaxy stellar mass function and the size-mass relation of late-type galaxies. We focus primarily on present-day subhaloes, but also examine the simulations at earlier times to explore evolutionary trends.

A summary of our results is as follows:

\begin{enumerate}

    \item The star-forming gas distribution of present-day subhaloes is typically flattened (i.e. low sphericity) along its minor axis. Flattening is most pronounced in subhaloes of $M_{\rm sub}\sim 10^{12.5}\Msun$, for which the median sphericity is $\tilde{S}_{\rm SF-gas}=0.1$. The distribution of star-forming gas sphericities is significantly narrower than that of stars and dark matter at all subhalo masses, but particularly for those of $M_{\rm sub}=10^{12-12.5}\Msun$, for which the interquartile ranges of star-forming gas, stars and DM are $0.06$, $0.15$ and $0.12$, respectively (Fig. \ref{fig:pdfs}). 
    
    \item Star-forming gas exhibits a diverse range of triaxiality parameters. Subhaloes of mass $M_{\rm sub}\sim10^{12-12.5}\Msun$ typically host oblate distributions consistent with classical gas discs, but in both low and high mass subhaloes, the distributions are more often prolate (Fig. \ref{fig:pdfs}). 
    
    \item Star-forming gas is less flattened at earlier epochs, for all subhalo masses examined, irrespective of whether one considers a sample selected in a similar fashion to the present-day sample, or considers the progenitors of the latter. Strongly flattened star-forming gas structures ($S\lesssim 0.2$) emerge only at $z\lesssim 2$, broadly coincident with the growth of the disc's scale length (Figs. \ref{fig:S_and_T_vs_mass} and \ref{fig:prog}).
    
    \item The shape parameters describing the morphology of star-forming gas are strongly and positively correlated with those describing the stellar morphology of the host galaxy, such that e.g. flattened gas structures are associated with flattened stellar structures (Fig. \ref{fig:shape_correlation}).
    
    \item The minor axis of the star-forming distribution preferentially aligns most closely with the minor axis of the (inner) DM halo. However, in prolate subhaloes $T_{\mathrm{DM}}(r < 30\mathrm{pkpc}) \gtrsim 0.7$, a significant fraction of galaxies have star-forming gas distributions whose minor axis most closely aligns with one of the other principal axes of the DM (Fig. \ref{fig:best_align}).
    
    \item Characterised by the angle between the minor axes of the respective components of subhaloes, star-forming gas tends to align with the DM (i.e. the alignment is stronger than random), but the alignment is weaker than is the case for stars and the DM. This is the case for both the 3D matter distributions (Fig. \ref{fig:3D_align}) and their projections on the sky (Fig. \ref{fig:2D_align}). The alignments are strongest when considering the inner DM halo, and in general the alignments are stronger for late-type galaxies.
    
    \item The alignment of the star-forming gas distribution with those of both the stars and the DM bound to its parent subhalo is typically weaker at early cosmic epochs (Fig. \ref{fig:3D_align_zdep}).
    
    \item The kinematic axis of star-forming gas aligns closely with its minor morphological axis, with most galaxies being aligned to better than $10^\circ$ at the present-day, and better than $6^\circ$ if only late-type galaxies are considered. The alignment is poorer at $z=1$, with these characteristic misalignment angles doubling (Fig. \ref{fig:sfg_minor_kin}).
    
    \item The more pronounced flattening of star-forming gas structures leads to them exhibiting a broader distribution of projected ellipticities than is the case for stellar structures, analogous to the differing ellipticity distributions of optical images of late-type and early-type galaxies. The ellipticity distribution of star-forming gas in EAGLE corresponds closely to that recovered from high-fidelity VLA radio continuum images of galaxies in the COSMOS field (Fig. \ref{fig:proj_view}). For a fixed subhalo sample, the `shape noise' of its star-forming gas is therefore systematically greater than that of its stars \ref{fig:shape_noise}).
    
\end{enumerate}

Our analyses reveal that the morphology of star-forming gas distributions, and their orientation with respect to the DM of their parent subhalo, are more complex than might be na\"ively assumed. This complexity is particularly relevant in the context of using extended star-forming gas distributions, which can be imaged in the radio continuum, to conduct weak lensing experiments. 

Forecasts for the outcomes of the next generation of the `megasurveys' require that very large cosmic volumes are modelled. The associated expense of including the baryonic component forces the use of empirical, analytic or semi-analytic models grafted onto treatments of the evolving cosmic dark matter distribution. By construction, such techniques do not capture the full complexity of the evolution of the baryonic component resulting from the diverse range of physical processes that influence galaxies, nor do they capture the `back reaction' of the baryons on the DM, and so can mask the importance of key systematic uncertainties. 

In the specific case of modelling the radio continuum sky, the most popular approach has been to couple observed source populations with either a Press-Schechter or $N$-body treatment of the evolving cosmic DM distribution \citep[see e.g.][]{wilman_08,bonaldi_19}. By construction, such models invoke no explicit connection between the properties of star-forming gas structures and their parent DM haloes, and often relate (or equate) the properties of the former to those of the host galaxy's stellar component. Our analyses highlight shortcomings of these approximations: the characteristic morphology of star-forming gas is a strong function of the mass of its host subhalo and, although the simulations indicate that it correlates strongly with the morphology of its associated stellar component, we find that the respective morphologies can differ significantly.

We also find that star-forming gas structures are imperfectly aligned with both the stellar and DM components of their host subhalo. Although the misalignment angle is generally small (particularly with respect to the stellar component), there is a long-tail to severe misalignments, and we find that the misalignment is most pronounced in early-type galaxies. We also find that the misalignment of the star-forming gas with the DM of its host subhalo becomes more pronounced if the outer halo is considered (for instance, if disabling the use of the $30\pkpc$ spherical aperture). Therefore, when constructing semi-empirical radio sky models based on $N$-body simulations, we caution against na\"ively orienting star-forming discs with the principle axes of the DM distribution.  

Our analyses also highlight that the shape noise of images of a fixed sample of galaxies seen in the radio continuum should be significantly greater than when seen in the optical. This follows naturally from the lower characteristic sphericity (or, alternatively, the greater flattening) of star-forming gas structures than their stellar counterparts. A systematic offset in shape noise was previously reported by \citet{tunbridge} following the examination of a relatively small sample of galaxies with high-fidelity radio and optical imaging. The corollary of this finding is that radio continuum weak lensing experiments will require a greater source density in order to obtain a signal-to-noise ratio equal to optical experiments. However, our analyses also corroborate the hypothesis that the use of the kinematic axis (revealed by ancillary 21 cm observations) affords an effective means of estimating the unsheared orientation of the minor axis, and thus mitigating the systematic uncertainty in radio weak lensing experiments.

An interesting consequence of the poorer alignment of star-forming gas structures with the DM of their host subhaloes than is the case for the stars - DM alignment, is that it implies that the intrinsic alignment signal may be less severe in radio weak lensing surveys than is the case for optical counterparts. In a follow-up paper, Hill et al. (in preparation), we examine the two key `intrinsic alignment' signals recoverable from radio continuum imaging, namely the orientation of star-forming gas distributions with respect to the directions to, and orientations of, the star-forming gas structures of its neighbouring galaxies.

\section*{Acknowledgements}
We gratefully thank Adrien Thob and Jonathan Davies for fruitful discussions and technical help. We thank the anonymous reviewer whose constructive suggestions improved both the content and presentation of this study. ADH is supported by an STFC doctoral studentship within the Liverpool Big Data Science Centre for Doctoral Training, hosted by Liverpool John Moores University and the University of Liverpool [ST/P006752/1]. RAC is a Royal Society University Research Fellow. This project has received funding from the European Research Council (ERC) under the European Union's Horizon 2020 research and innovation programme (grant agreement No 769130). The study made use of high performance computing facilities at Liverpool John Moores University, partly funded by the Royal Society and LJMU's Faculty of Engineering and Technology, and the DiRAC Data Centric system at Durham University, operated by the Institute for Computational Cosmology on behalf of the STFC DiRAC HPC Facility (www.dirac.ac.uk). This equipment was funded by BIS National E-infrastructure capital grant ST/K00042X/1, STFC capital grants ST/H008519/1 and ST/K00087X/1, STFC DiRAC Operations grant ST/K003267/1 and Durham University. DiRAC is part of the National E-Infrastructure.

\section*{Data Availability}

Particle data, and derived data products from the simulations have been released to the community as detailed by \citet{mcalpine16}. Further derived data used in this article will be shared on reasonable request to the corresponding author.


\bibliographystyle{mnras} 
\bibliography{lib1} 


\begin{appendix}

\section{Numerical Convergence}\label{sec:num_conv}

\begin{figure*}
\centering
     \includegraphics[width = 0.95\textwidth]{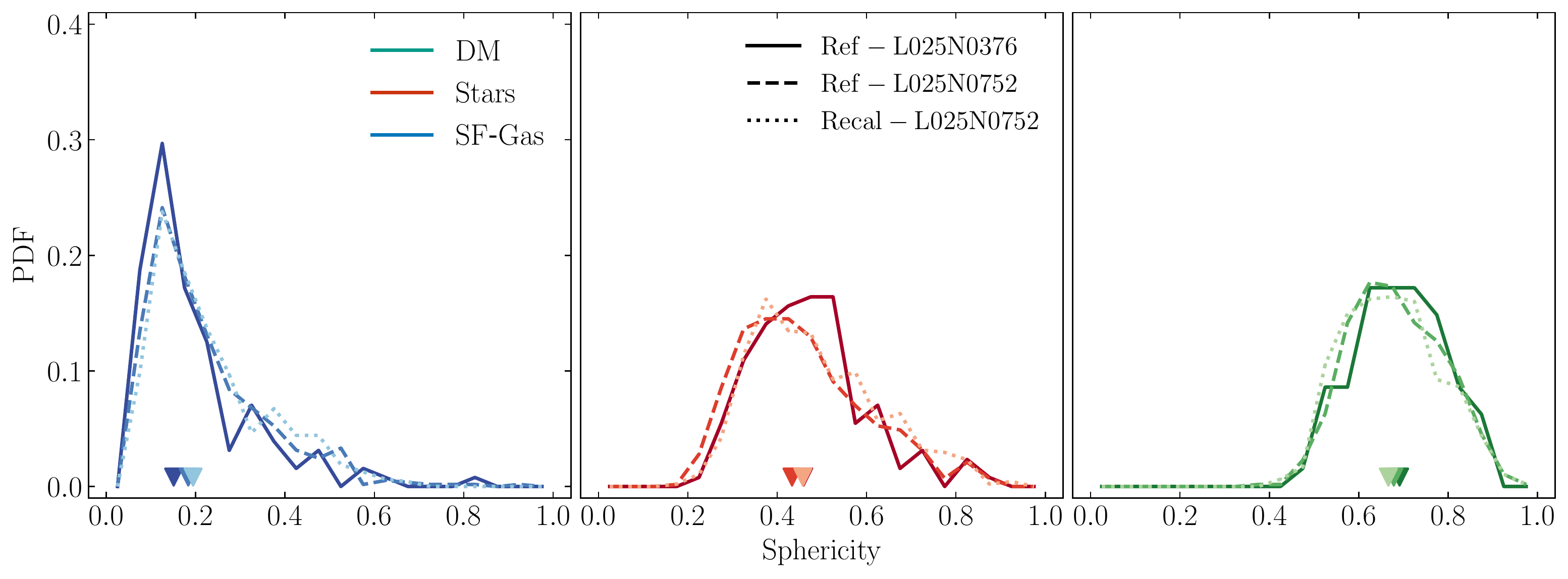}
\caption{Probability distribution function of the sphericity parameter of the star-forming gas (left-hand panel), stars (centre) and DM (right-hand panel) of present-day subhaloes drawn from the Ref-L025N0376 (solid dark-coloured curve), Ref-L025N0752 (dashed medium), and Recal-L025N0752 (dotted light) simulations. Down arrows denote the median sphericity of the distribution of each simulation. Comparison of Ref-L025N0376 with Ref-L025N0752 and Recal-L025N0752 affords simple tests of, respectively, the strong and weak convergence behaviour of the star-forming gas sphericity.}\label{fig:pdfs_conv}
\end{figure*}

In this section we examine the influence of the numerical resolution of the EAGLE simulations on the recovered sphericity of the star-forming gas, stars and DM comprising subhaloes. We follow \citet{schaye15} and adopt the terms `strong convergence' and `weak convergence', where the former denotes a comparison at different resolutions of a fixed physical model, and the latter denotes a comparison at different resolutions of two models calibrated to recover the same observables. We use three $L=25\cMpc$ simulations introduced by \citet{schaye15}: Ref-L025N0376, which is identical to the flagship Ref-L100N1504 simulation with the exception of the boxsize; Ref-L025N0752, which adopts the same Reference physical model but has a factor of 8 more particles each of both baryons and DM; and Recal-L025N0752 which also adopts values for subgrid parameters governing stellar and AGN feedback that have been recalibrated to improve the match to the galaxy stellar mass function and galaxy sizes at this higher resolution. Comparison of  Ref-L025N0376 with Ref-L025N0752 and Recal-L025N0752 thus affords simple tests of, respectively, the strong and weak convergence behaviour.

Fig. \ref{fig:pdfs_conv} shows the probability distribution functions of the sphericities of the star-forming gas (left-hand panel), stars (centre) and DM (right-hand panel) for each of the three $L=25\cMpc$ simulations. The subhaloes shown are selected according to the standard sampling criteria outlined in the Section \ref{sec:sample}, irrespective of the resolution of the simulation. Down arrows denote the median sphericity of the distribution of each simulation. Inspection shows that the distributions are not strongly influenced by the change in resolution. The median values of the sphericity of the three matter components in the Ref-L025N0376 simulation are 0.15, 0.50 and 0.69 for the star-forming gas, stars and DM, respectively. As can be clearly seen from the figure, when moving to the high resolution simulations, the shift in median values is much smaller than the associated interquartile ranges of the Ref-L025N0376 simulation (IQR = $0.14$, $0.15$, $0.14$ for the three components, respectively). Although not shown here, we recover similar behaviour when focusing on the triaxiality parameter.

\section{Influence of subgrid ISM treatments}\label{sec:eos_conv}

\begin{figure}
\centering
     \includegraphics[width = 0.45\textwidth]{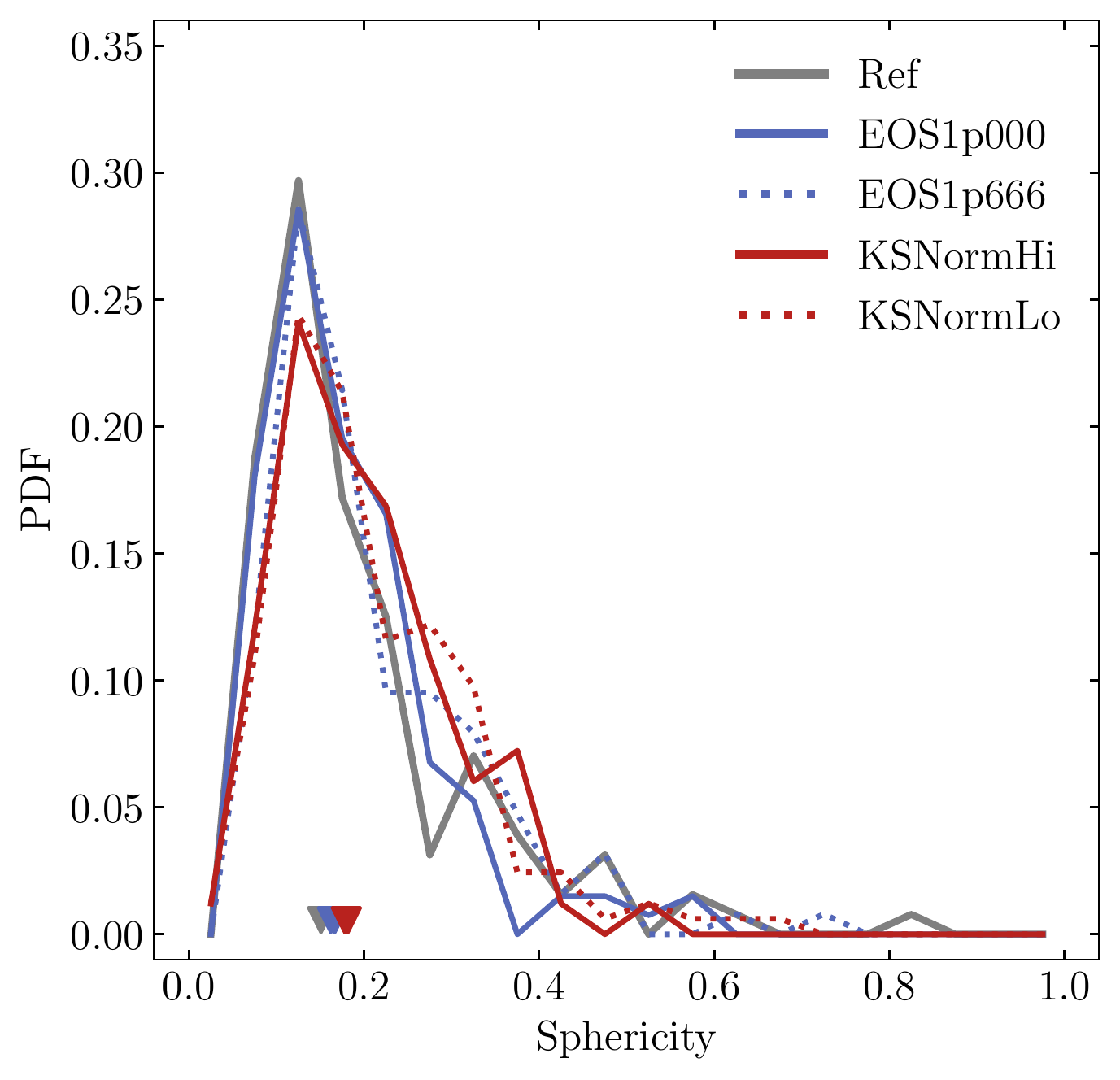}
    \caption{Probability distribution function of the sphericity parameter of the star-forming gas of present-day subhaloes drawn from the Ref-L025N0376 simulation (black curve) and two pairs of simulations that incorporate variations of the reference model:  pair with alternative equation of state slopes (EOS1p00, solid blue; and EOS1p666, dotted blue) and with normalisations of the star formation law adjusted by $\pm 0.5$ dex (KSNormHi, solid red; KSNormLo, dotted red). Comparison of these runs with the reference model indicates the influence of the subgrid ISM model on the sphericity of star-forming gas distributions.}\label{fig:eos_conv}
\end{figure}

In this section we examine the sensitivity of star-forming gas morphologies to aspects of EAGLE's subgrid models that in principle influence the structure of interstellar gas directly, namely the form of the temperature floor equation of state and the star formation law. To achieve this, we compare the Ref-L025N0376 simulation with two pairs of complementary L025N0376 simulations. The first pair, introduced by \citet{crain15}, varies the slope of the equations of state from the reference value of $\gamma_{\rm eos}=4/3$ with different slopes, to adopt isothermal ($\gamma_{\rm eos}=1$) and adiabatic ($\gamma_{\rm eos}=5/3$) equations of state. \citet{schaye_DV} used simulations of idealised discs to show that a stiffer equation of state generally leads to smoother star-forming gas distributions with a larger scale height. \citet{crain15} showed that in EAGLE, a stiffer equation of state also suppresses accretion onto the central BH in massive galaxies. The second pair, introduced by \citet{crain17}, varies the normalisation of the Kennicutt-Schmidt law \citep[the variable $A$ in equation 1 of][]{schaye15} from its fiducial value of $1.515\times10^{-4}\Msunyrsqkpc$ by $\pm 0.5$ dex. \citet{crain17} showed that increasing (decreasing) this parameter tends to decrease (increase) the mass of cold gas associated with galaxies, since it governs the mass of gas that is required to maintain a balance between the gas infall rate and the outflow rate due to ejective feedback.

Fig. \ref{fig:eos_conv} shows probability distribution function of the sphericity star-forming gas for the reference model (solid black curve) and the simulations with differing equations of state ($\gamma_{\rm eos}=1$, solid blue; $\gamma_{\rm eos}=5/3$, dotted blue), and with higher (solid red) and lower (dotted red) normalisations of the star formation law with respect to the reference model. The subhaloes shown are selected according to the standard sampling criteria outlined in the Section \ref{sec:sample}. Down arrows denote the median sphericity of the distribution of each simulation. Inspection reveals that the distributions are not strongly influenced by changes to the subgrid modelling of the ISM. The median value of the sphericity of the star-forming gas in the Ref-L025N0376 simulation is 0.15. As can be clearly seen from the figure, the median sphericity in the three variation simulations shifts by $< 0.05$ with respect to the reference simulation, a value that is much smaller than the interquartile range of the reference case. Although not shown here, we recover similar behaviour when focusing on the triaxiality parameter.  

\section{The influence of particle sampling on shape characterisation}\label{sec:min_part}

\begin{figure}
     \includegraphics[width = \columnwidth]{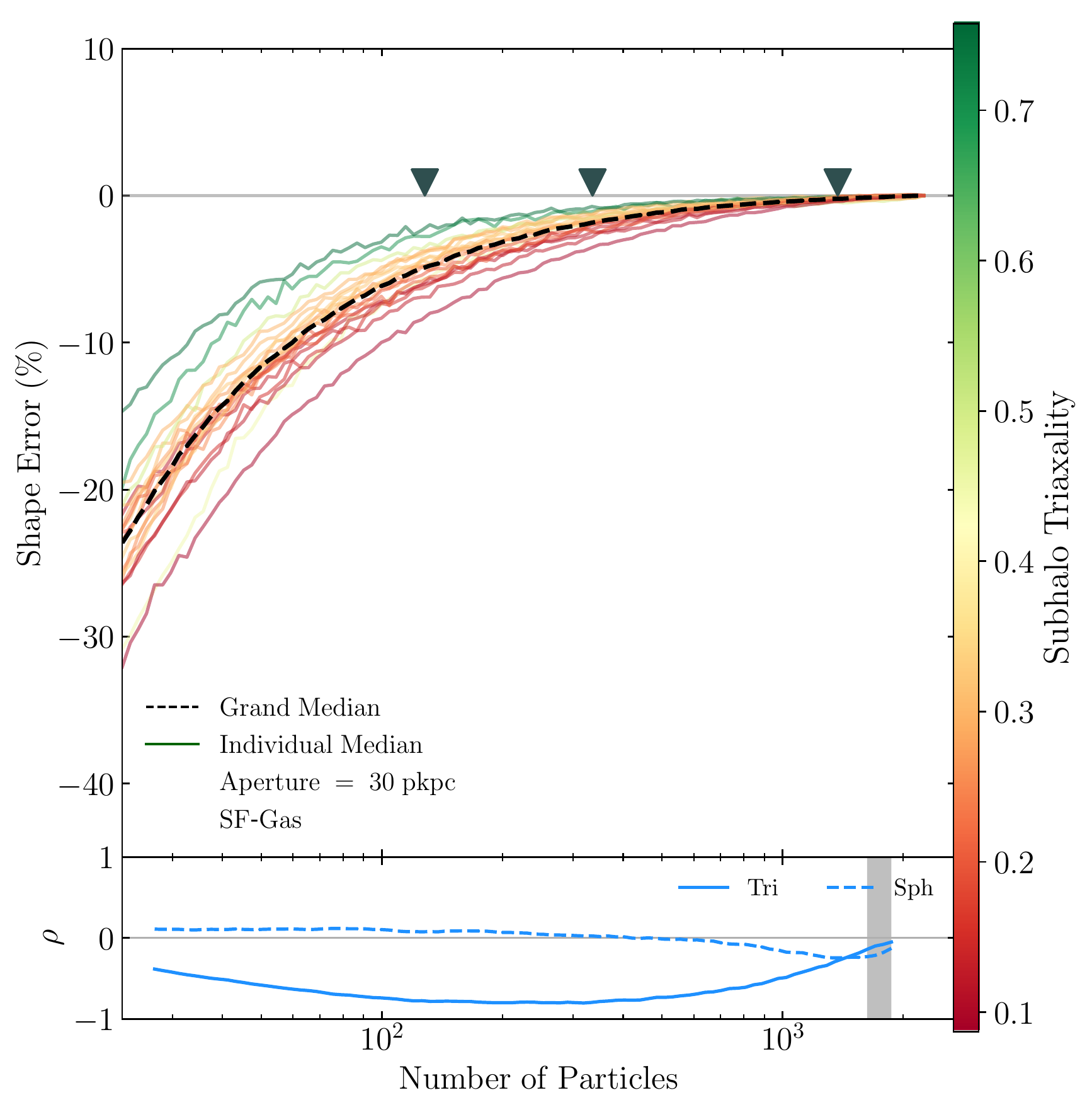}     
\caption{The sphericity shape error recovered as a function of the degree of particle subsampling for star-forming gas in 20 present-day subhaloes with dynamical mass similar to that of the Milky Way. Down arrows correspond to the 10th, 50th and 90th percentile values of the number of star-forming gas particles within the subhalo sample. Curves show the median shape error recovered from $10^5$ random subsamplings of the true star-forming gas particle distribution, and are coloured by the latter's true triaxiality. The dashed curve shows the median recovered by aggregating measurements from all 20 subhaloes. The subpanel shows the running Spearman rank correlation coefficients relating the shape error to the true value of the triaxiality (solid curve) and sphericity (dashed curve) of the star-forming gas.}
\label{fig:shape_convergence}
\end{figure}

The morphological characterisation of structures defined by particle distributions is unavoidably influenced by sampling error. It is therefore crucial to establish the reliability of such characterisations as a function of particle number. A common methodology is to realise a mass distribution of a known analytic form with a particle distribution, and assess the deviation of the recovered shape from the input shape as the distribution is progressively subsampled \citep[see e.g. Appendix A2 of][]{vel15a}. We adopt a similar approach here but, since star-forming gas distributions are not readily characterised by a simple analytic form, we instead draw 20 central subhaloes from the sample described in Section \ref{sec:sample}, with dynamical mass comparable to that of the Milky Way ($M_{\rm sub} \simeq 10^{12.0-12.5}\mathrm{M_{\odot}}$). We compute their `true' shape parameters by applying the algorithm defined in Section \ref{sec:redit} using all star-forming gas particles ($N_{\rm part} \simeq 2000$). We then progressively subsample the particle distribution to lower $N_{\rm part}$, generating $10^5$ realisations at each value of $N_{\rm part}$, and recompute the shape parameters.

Fig.~\ref{fig:shape_convergence} shows the median of the relative error on the sphericity of the star-forming gas distribution recovered from the $10^5$ subsamplings of the particle distribution as a function of $N_{\rm part}$. The curves are coloured by the `true' value of the triaxiality parameter of the subhalo's star-forming gas. The dashed black curve shows the `grand median' recovered by aggregating the measurements from all 20 subhaloes. Down arrows show the 10th, 50th and 90th percentile values of the number of star-forming gas particles within the subhalo sample. \citet{vel15a} noted that poor particle sampling leads to a systematic underestimate of the sphericity parameter of the DM; we find this is also the case for the star-forming gas. A shape error of less than 10 percent typically requires at least $N_{\rm part}= 100$, hence we adopt this threshold as our lower limit for our sample selection.

Inspection of the curves for the individual subhaloes indicates that this value is sensitive to the triaxiality of the structure, with accurate recovery of the sphericity requiring fewer particles in prolate ($T>0.5$) distributions. As shown in Fig.~\ref{fig:S_and_T_vs_mass}, the star-forming gas of low-mass subhaloes is preferentially prolate, hence a minimum of $N_{\rm part}= 100$ can be considered a conservative choice. For completeness, the subpanel shows the `running' value of the Spearman rank coefficient recovered from $N_{\rm part}$-ordered subsamples, of the correlation between the \textit{absolute} shape error and the true shape parameters, $T$ (solid curve) and $S$ (dashed curve). The solid curve highlights that a negative correlation between the shape error on sphericity and the true triaxiality persists to over $1000$ particles. The dashed curve indicates that there is a very mild positive correlation of the relative shape error on sphericity with the true input sphericity. 

\section{Analytic fits to the misalignment angle distributions} \label{sec:fits}

\begin{figure*}
\centering
     \includegraphics[width = 0.94\textwidth]{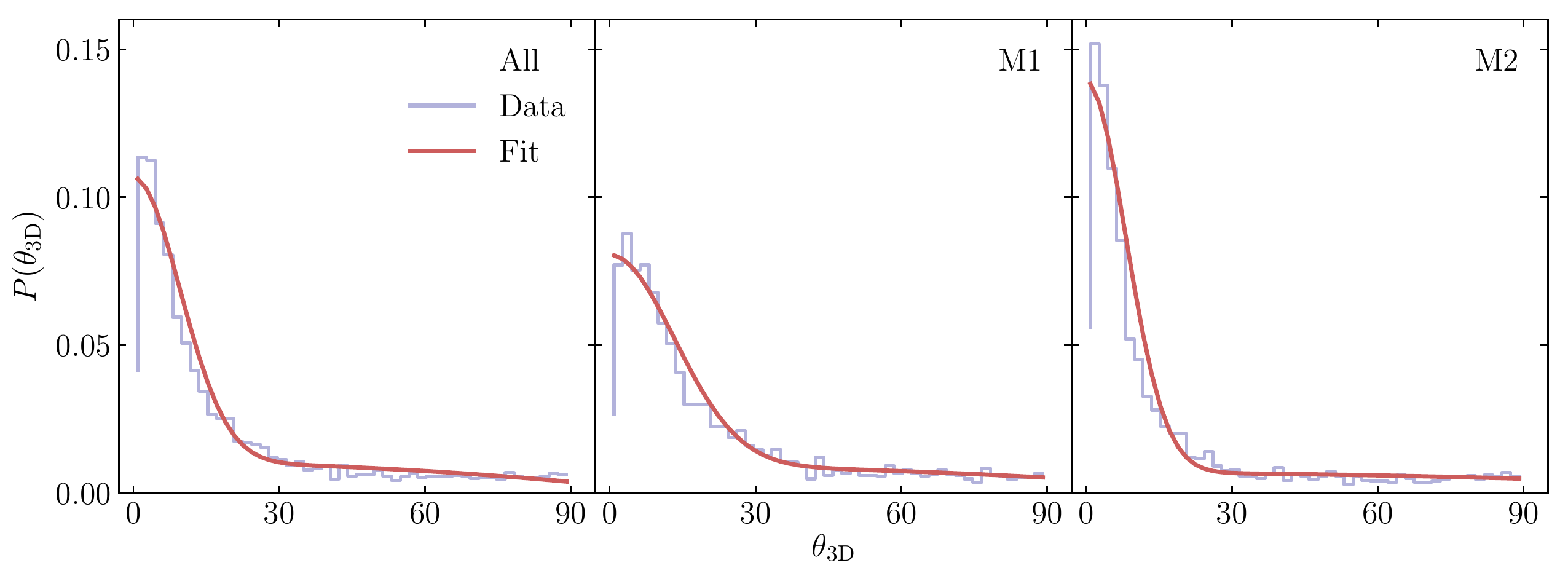}
\caption{Probability distribution functions $P(\theta_{\mathrm{3D}})$, where $\theta_{\mathrm{3D}}$ is the angle between the morphological minor axes of stars and DM within the sample of subhaloes. A fiducial aperture of $30\pkpc$ is imposed. The faded step functions show the raw histograms, while the smooth lines are their respective analytic fits described by equation~\ref{eq:fitting_any_D}. Panels correspond to different mass bins: the full sample (left-hand panel), $M_{\mathrm{sub}} \leq 10^{11.5}\mathrm{M_{\odot}}$ (middle) and $10^{11.5}\mathrm{M_{\odot}}\leq M_{\mathrm{sub}}<10^{13}\mathrm{M_{\odot}}$ (right-hand panel). During fitting, errors in the $y$-axis were taken to be the 1$\sigma$ Poisson errors.}\label{fig:fitting}
\end{figure*}

We provide fitting functions to the distribution of internal misalignment angles between star-forming gas and DM for present-day subhaloes in three mass bins from the EAGLE Ref-L100N1504 simulation, in both 2- and 3D. The fits enable users of $N$-body simulations to populate subhaloes with galaxies oriented with respect to the minor axis of the subhalo in a realistic fashion. We fit to $P({\theta})$ using the following functional form:
\begin{equation}
    \mathcal{M}(\theta) = C\mathrm{exp}\Big(-\frac{\theta^{2}}{2\sigma_{1}^{2}}\Big) + D\mathrm{exp}\Big(-\frac{\theta^{2}}{2\sigma_{2}^{2}}\Big) + E\label{eq:fitting_any_D},
\end{equation}
where $C, D, \sigma_{1}^{2}, \sigma_{2}^{2}, E$ are the free parameters, and $\theta$ is the misalignment angle. The same form was used by \citet{vel15a} to fit to the misalignment angle of stars and DM in projection. We calcuate the best fit parameters with the Python package \textsc{scipy.optimize.curve\_fit}, using 1$\sigma$ Poisson errors. 

The best fit parameters are quoted in Tables~\ref{tab:align_fit}. Parameters are recovered for the misalignment angles in both the cases of i) applying our fiducial aperture to the initial step of the iterative algorithm, and ii) applying no initial aperture, i.e. considering all particles bound to the subhalo. In addition to presenting best fit parameters for all subhaloes in our sample (`All'), we provide fits to subsamples `M1' and `M2', which are subject to the additional criteria $\mathrm{log_{10}}M_{\mathrm{sub}}[\mathrm{M_{\odot}}]\leq 11.5$ (M1) and $11.5 < \mathrm{log_{10}}M_{\mathrm{sub}}[\mathrm{M_{\odot}}]\leq 13$ (M2). This is motivated by two factors. Firstly, below $M_{\rm sub} = 10^{11.5}\Msun$ our selection criteria result in significant incompleteness. Secondly, the misalignment of the minor axes of the star-forming gas and the DM components becomes large for $M_{\rm sub} > 10^{13}\mathrm{M_{\odot}}$ (see discussion in Section~\ref{sec:morphological_alignments}), severely degrading the value of the fits. The best fits for the 3D fiducial aperture case are shown in Fig. \ref{fig:fitting}.

We find that the fitting is able to recover the profile of the input distribution fairly successfully. As an example we find the percentage difference in the retrieved median as compared with the input distribution to be $(1.7, 0.14, 0.51)$ percent for the three cases displayed in the panels of Fig~\ref{fig:fitting}, while for the standard deviation this becomes $(0.72, 1.04, 1.16)$ percent. For the no aperture version of these cases we find errors of $(3.1, 2.2, 2.6)$ percent for the median and $(1.6, 1.8, 1.7)$ percent for the standard deviation. When no aperture is applied, we find that the errors in the 2D fittings are comparable to the 3D case. However with the fiducial aperture the errors are noticeably larger for the 2D case, the largest being $\sim10$ percent for the median and standard deviation of the M2 bin. Twelve figures comprising all variations displayed in Table~\ref{tab:align_fit} (two dimensions $\times$ two apertures $\times$ three mass bins) in the style of Fig~\ref{fig:fitting} may be found at the author's website\footnote{\url{www.astro.ljmu.ac.uk/~ariahill/}}.

\begin{table*}
\centering
\begin{tabular}{l | c  c  c c c c|c c c c  c  c}
&&&2D&&&&&3D\\
\cmidrule(lr){2-6}\cmidrule(lr){7-11}
\textbf{Aperture} \& \textit{Mass-Bin}  & $C$ & $D$ & $E$ & $\sigma_{1}$ & $\sigma_{2}$& $C$ & $D$ & $E$ & $\sigma_{1}$ & $\sigma_{2}$ \\
\cmidrule(lr){1-11}
\textbf{Fiducial}\\
\textit{All} & 24.0 & 8.33 & 0.0199 & 0.0507 & 0.00717 & 5130.0 & 9.62 & 43.2 & 0.0962 & -43.2 \\
\textit{M1} & -27.0 & 9.21 & 0.0205 & 0.042 & 0.00688 & -13.3 & 3050.0 & 0.0714 & 9.06 & -9.05   \\
\textit{M2} & 9.32 & -9.32 & -218.0 & 218.0 & 0.0107 & -8.13 & 1440.0 & 0.132 & 1.03 & -1.03  \\
\cmidrule(lr){2-6}\cmidrule(lr){7-11}
\textbf{No Aperture}\\
\textit{All} & -17.4 & 17.4 & -87.3 & 87.3 & 0.0122 & 29.3 & -29.3 & -157.0 & 157.0 & 0.0125 \\
\textit{M1} & -17.7 & 17.7 & -71.7 & 71.7 & 0.0124 & 31.2 & -31.2 & -181.0 & 181.0 & 0.013 \\
\textit{M2} &  -17.3 & 17.3 & -79.8 & 79.8 & 0.0124 & 26.9 & -26.9 & -114.0 & 114.0 & 0.0121 \\

\cmidrule(lr){1-11}
\end{tabular}
\caption{Best fitting parameters for equation~\ref{eq:fitting_any_D}, used to fit the probability distribution functions of the intrinsic 3D and projected 2D misalignment angle between star-forming gas and DM within present-day subhaloes of three mass bins (denoted by italics). Parameters are provided for the angles recovered using our fiducial initial aperture for the iterative reduced inertia tensor, and for no aperture (denoted by font weight).}
\label{tab:align_fit}
\end{table*}

\end{appendix}
\label{lastpage}

\end{document}